\documentclass{pasa}
\usepackage{color}
\newcommand{\nar}{NewAR}
\newcommand{\mnras}{MNRAS}
\newcommand{\msun}{~\mathrm{M}_{\odot}}
\newcommand{\zsun}{~$\rm{Z}_{\odot}$ }

\newcommand{\adb}[1]{\textcolor{blue}{ #1}} 

\title[First SMBHs]{On the formation of the first quasars}
\author[Valiante et al.]{Rosa Valiante$^1$\thanks{rosa.valiante@oa-roma.inaf.it}, Bhaskar Agarwal$^{2,3}$, Melanie Habouzit$^4$ \and Edwige Pezzulli$^{1,5}$\\
\affil{$^1$INAF/Osservatorio Astronomico di Roma, Via di Frascati 33, 00040 Monte Porzio Catone, Italy}%
\affil{$^2$Universit\"{a}t Heidelberg, Zentrum f\"{u}r Astronomie, Institut f\"{u}r Theoretische Astrophysik, \\   
Albert-Ueberle-Stra{\ss}e 2, 69120 Heidelberg, Germany}
\affil{$^3$Department of Astronomy, 52 Hillhouse Avenue, Steinbach Hall, Yale University, New Haven, CT 06511, USA}%
\affil{$^4$Center for Computational Astrophysics (CCA), Flatiron
Institute, 162 5th Ave New York}%
\affil{$^5$Dipartimento di Fisica, Universit{\'a} di Roma  ``La Sapienza'', P.le Aldo Moro 2, 00185, Roma, Italy}
}
\jid{PASA}
\doi{10.1017/pas.\the\year.xxx}
\jyear{\the\year}

\usepackage[authoryear]{natbib}
\bibpunct{(}{)}{;}{a}{}{,}
\usepackage{aas_macros}
\usepackage{hyperref} 
\hypersetup{colorlinks,citecolor=blue,linkcolor=blue,urlcolor=blue}

\begin{document}%
\begin{abstract}
Observations of the most luminous quasars at redshift $z>6$ reveal the existence of 
numerous supermasssive black holes ($>10^9\msun$) already in place about 
twelve billion years ago. 
In addition, the interstellar medium of the galaxies hosting
these black holes are observed to be chemically mature systems, with metallicities 
($Z>Z_\odot$) and dust masses ($>10^8 \rm M_\odot$) similar to that of more evolved, 
local galaxies.
The connection between the rapid growth of the first supermassive black holes and 
the fast chemical evolution of the host galaxy is one of the most puzzling issues 
for theoretical models. 
{Here we review state-of-the-art theoretical models that focus on this problem with particular emphasis on the conditions that lead to the formation of quasar seeds and their subsequent evolution at $z\geq 6$.}
\end{abstract}
\maketitle%
\section{INTRODUCTION}
\label{sec:intro}
Up to $\sim$ 40 {supermassive} black holes (SMBHs) of $>10^9\msun$ {have been observed till date, which are believed to power the optically bright quasars ($>10^{47}$ erg s$^{-1}$) at $z>5$ (e.g. \citealt{Mortlock11, Wu15})}.
{How these BHs formed in a relatively short time scale, already 12 Gyr ago in the \adb{early} Universe ($\lesssim 700-800$ Myr; e.g. \citeauthor{Fan01} \citeyear{Fan01, Fan04}, 
\citeauthor{deRosa11} \citeyear{deRosa11, deRosa14}) is still an open question (e.g. \citealt{Volonteri10, Natarajan11}).

Luminous (optically selected) quasars at high redshift, thus {offer} the 
most direct constraint on the evolution of the first SMBHs and {serve} a unique 
laboratory to study the earliest phases of galaxy formation and evolution 
as well as the properties of the early Universe.
In the left panel of Fig.~\ref{fig:fig1} we show a collection of high redshift ($z>3$) SMBHs {reported to date. Note that at $z>6$}, they are already as massive as the BHs observed at lower redshifts ($z=3-5$) and in the local Universe (see e.g. \citealt{Sani11}, \citealt*{KHo13}).
\begin{figure*}
\begin{center}
\includegraphics[width=8cm]{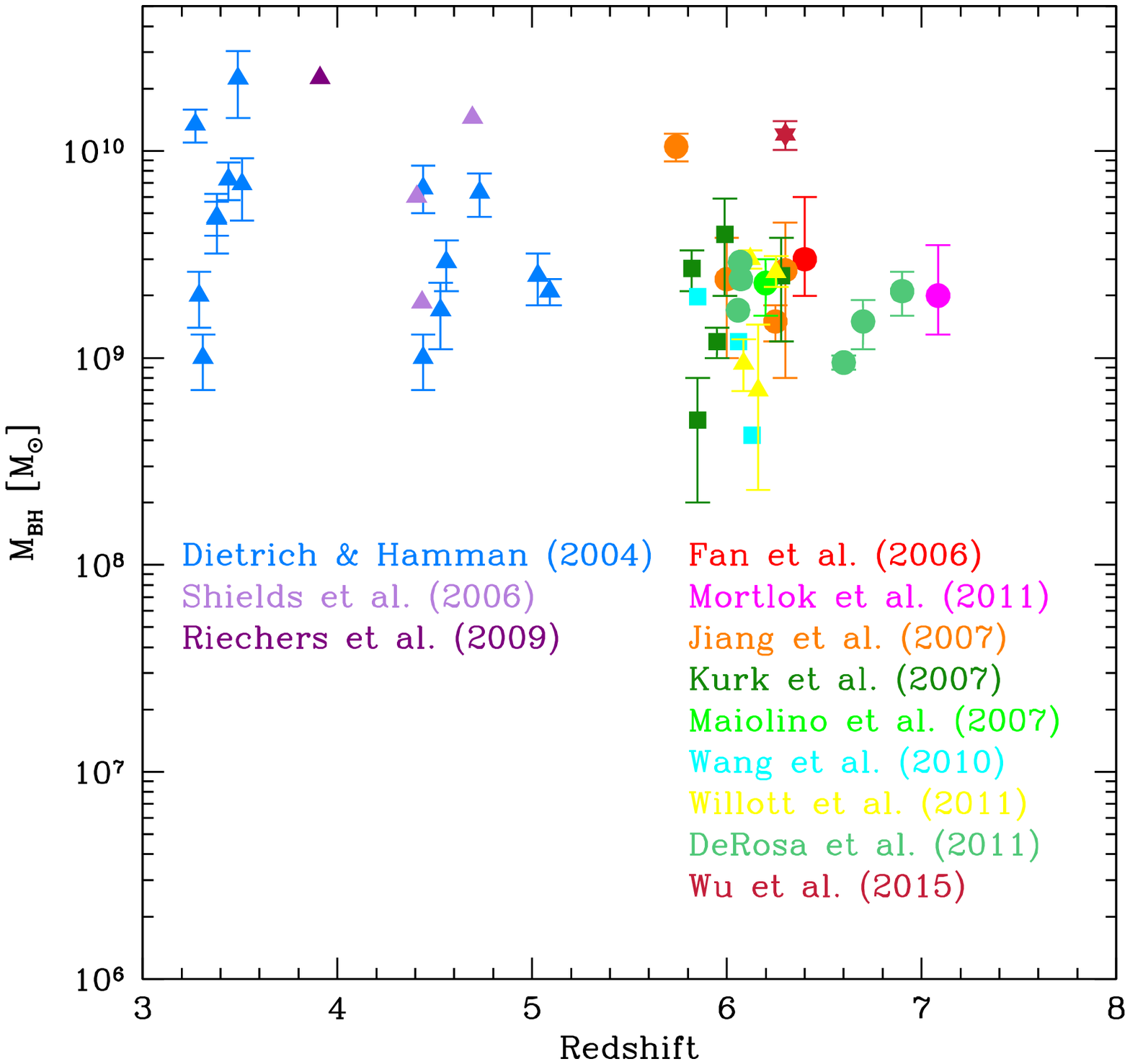}
\includegraphics[width=8cm]{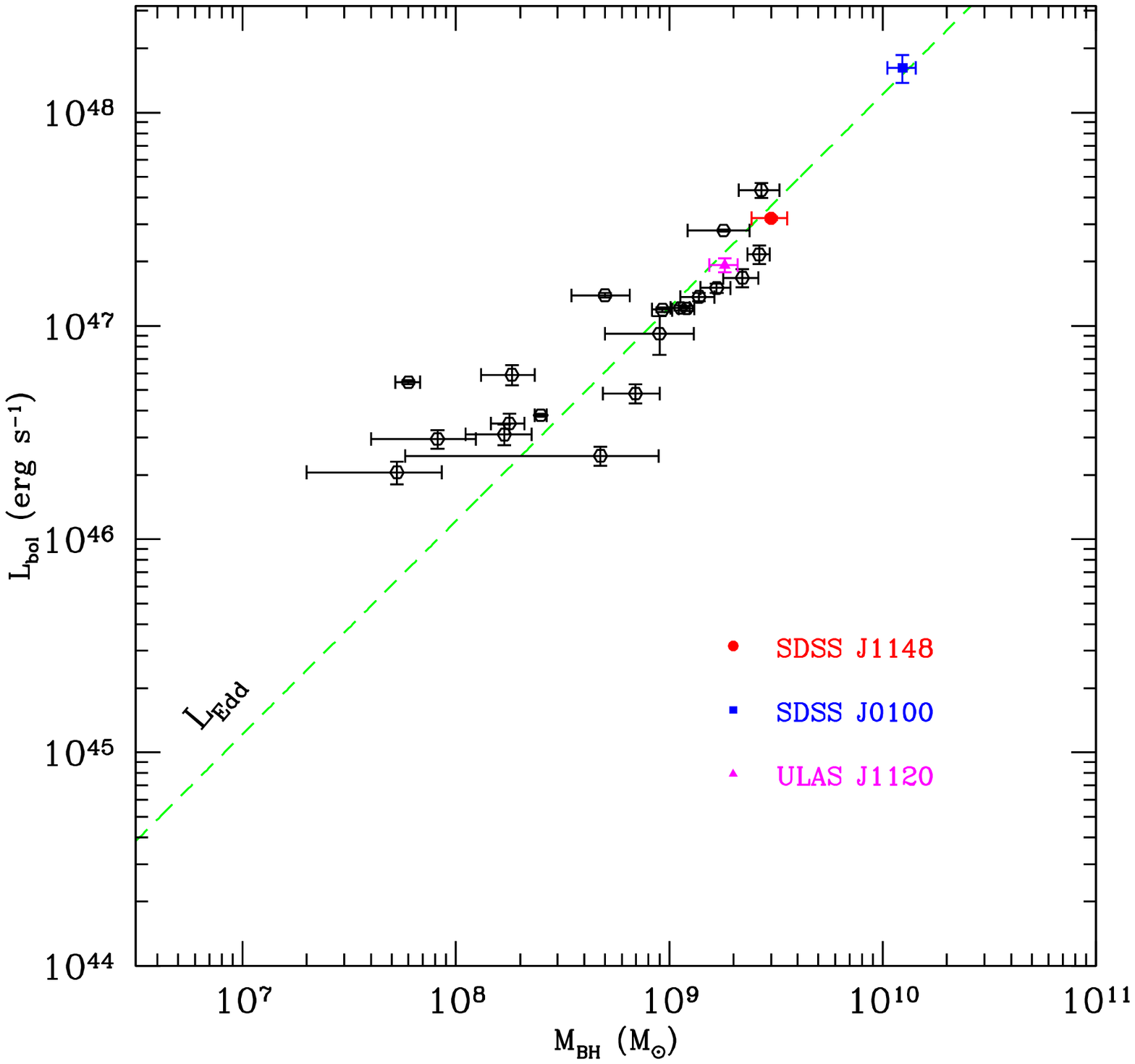}
\caption{
\textit{Left panel}: Black hole mass as a function of redshift in $z>3$ galaxies. References to the data are labeled and color coded in the figure. 
\textit{Right panel}: Bolometric luminosity as a function of the black hole mass for $z\sim6$ quasars. Blue square, red circle and magenta triangle represent quasars J0100, J1148 and J1120, respectively. Black empty data points are from the collection of high-z quasars by \cite{Wu15}. The green dashed line show Eddington luminosity (Courtesy of F. Wang and X.B. Wu).}\label{fig:fig1}
\end{center}
\end{figure*}
{The two noteworthy} record holders are ULAS J1120+0641 (J1120) and SDSS J0100+2802 (J0100), hosting the most distant ($z\sim 7.1$, \citealt{Mortlock11}), and the most massive ($1.2\times 10^{10} \rm M_\odot$, \citealt{Wu15}) SMBHs ever observed respectively.

In the right panel of Fig.~\ref{fig:fig1} we show the bolometric luminosity as a function of the BH mass for the collection of $z\geq 6$ quasars presented by \citet{Wu15}.\\ 
The nuclei of these objects are actively accreting massive BHs,
shining close to or above the Eddington luminosity (green dashed line). 
Colored points show three of the most interesting objects observed to date: 
the two record holders introduced above, J1120 (magenta triangle) and J0100 
(blue square) and quasar SDSS J1148+5251 (red circle, hereafter J1148) which 
is one of the best studied quasar, discovered at $z=6.4$ \citep{Fan01}.
As it can be seen from the figure, J0100 is the most luminous quasar known at 
$z>6$, with bolometric luminosity $L_{\rm Bol}=L_{\rm Edd}\sim 4\times 10^{14}$ 
L$_\odot$\citep{Wu15}, {making it 4 times brighter than J1148 (red circle), and 7 times brighter than J1120 (magenta triangle).}

The existence of these active and massive BHs close to the reionization epoch 
when the Universe was younger than $\sim 1$ Gyr, triggered a number of 
theoretical studies and deep, systematic observational campaigns that aimed at shedding light on their origin (e.g. \citealt{Willott10, Wu15, Venemans15, Banados16} and references therein). 
This manuscript is part of a series of reviews on high redshift BHs and 
it is intended to present state-of-the-art theoretical models for the 
formation and evolution of high redshift SMBHs and their host galaxies.

\section{OPEN QUESTIONS}
\label{sec:questions}
Currently, many efforts are devoted to explaining how and when the first 
BHs and their host galaxies form. 
Here we briefly discuss the mostly debated issues related to the discovery of 
distant quasars and their observed properties (see \citealt{Gallerani17} for a 
recent review on the first quasars observed physical properties). 

\subsection*{{How and when did the $z>6$ SMBHs form \& the nature of 
their progenitors}}

{The formation mechanism and properties of the first seed BHs are the subject of several studies which focus on three distinct scenarios (see e.g. }
\citealt{Volonteri10,Natarajan11, LatifFerrara16} for complete reviews):
\begin{enumerate}
\item light seeds, namely stellar mass BHs ($\sim 100\msun$) 
formed as remnants of Population III (Pop~III) stars 
\item intermediate mass, $10^3-10^4\msun$, BHs arising from stars 
and stellar-mass BHs collisions in dense clusters 
\item heavy seeds, forming in $T_{\rm vir}\geq 10^4 K$ halos, exposed 
to an intense $\rm H_2$ photo-dissociating {ultra--violet (UV) flux}, via direct collapse (DC) of low metallicity gas clouds into $10^4-10^6\msun$ BHs.
\end{enumerate}

Another debated issue {is related to} the BH growth mechanism required to explain
$z>6$ SMBHs.
Different studies suggest that BHs may evolve via uninterrupted gas accretion at the Eddington rate and/or episodic super-Eddington accretion phases, to grow up to 
billion solar masses, especially in the case of light seeds. 
We refer the interested reader to reviews by \citet{Volonteri10, Natarajan11},
\citet*{VolonteriBellovary12}, \citet{Volonteri15b}, \citet*{LatifFerrara16}, 
\citet*{JohnsonHaardt16} and references therein for details on the 
first seed BHs formation and feeding mechanisms.

The seeds of the first SMBHs are still elusive even to the most {sensitive 
instruments that exist today, thus preventing us from putting observational constraints on their nature.
A good example is the bright} Ly$\rm \alpha$ emitter CR7 observed at $z\sim 6.6$
 \citep{Matthee15, Sobral15,Bowler16} where both: a stellar \citep{Sobral15, Visbal16, Dijkstra16}; and a direct collapse 
\citep{Pallottini15, Hartwig15, Agarwal16b, Smith16, Smidt16} origin has been suggested for its metal poor component.

Although the observational signatures of seed BHs still remain unexplored, 
\citet{Pacucci16} suggest a promising method to search for DCBH candidates in 
deep multi-wavelength surveys, based on photometric observations. 
By modeling the spectral energy distribution and colors of objects selected 
from the CANDELS/GOODS-S field catalogs \citep{Guo13} they identify two 
X-ray detected faint active galactic nuclei (AGN), 33160 and 29323 \citep{Giallongo15} 
as DCBHs prototypes at $z\sim 6$ and $\sim 9.7$, respectively. 

\subsection*{What are the properties of high-z SMBHs hosts?}
{High--z quasars are found to reside in over--dense environments (e.g. \citealt{Morselli14}), in galaxies that are 
chemically evolved, metal and dust-rich}. The metallicity of quasar host galaxies is quite difficult to trace.
Constraints on the gas-phase elemental abundances in their interstellar medium
(ISM) come from the detection of emission line ratios in broad- and the 
narrow-line regions (BLRs and NLRs, respectively). 

Although BLRs are representative of a small fraction of the gas content, 
concentrated within the central region ($10^4\msun$ on parsec scales, 
close to the AGN), the observed emission line ratios, such as 
FeII/MgII (e.g. \citealt{Barth03}), NV/CIV (e.g. \citealt{Pentericci02}), 
(Si IV+OIV)/CIV \citep{Nagao06, Juarez09}, and metal lines like CII and 
OI \citep[e.g.][]{Maiolino05, Becker06} trace up to $\sim 7$ Z$_\odot$ 
metallicities \citep{Nagao06, Juarez09} suggesting 
a fast evolution of the ISM chemical properties. 

A better proxy {of} the host galaxy ISM metallicity,
on larger scales (comparable to the host galaxy size), is 
provided by NLRs. 
A mean gas-phase metallicity $\rm Z_{NLR}=1.32^{+0.25}_{-0.22} Z_\odot$
is inferred from CIV/He II and C III/C IV flux ratios in quasar,
with no significant evolution up to $z\sim4$ \citep{Nagao06, Matsuoka09}.
{Such super-solar metallicities are a reminiscence of the star formation history (SFH) of the system (see e.g. \citealt{Matsuoka09} and references therein) and can serve as a lower limit for the $z\sim 6$ quasar host galaxies.}

Constraints on the dust content come from the observations of far-infrared 
(FIR) and sub-millimeter (sub-mm) continuum radiation.
The observed $\geq 10^{13}$ L$_\odot$ quasar FIR luminosities are 
consistent with emission from warm dust (30-60 K) with masses 
$>10^8\msun$ \citep{Bertoldi03, Priddey03, Robson04, Beelen06, 
Wang08, V11, V14, Michalowski10}.
From the same FIR luminosities, high star formation rates (SFRs), 
$\geq 1000$ M$_\odot/$yr, can be inferred, suggesting that a large fraction 
of these systems has ongoing, highly efficient, star-formation 
activity (see e.g. Table 1 in \citealt{V14} and references 
therein)\footnote{Note that the SFR is usually inferred using the FIR Luminosity-SFR scaling relation
\citep{Kennicutt98} which relies on the assumption that all FIR radiation comes from dust heated by stellar optical-UV 
emission. A factor of $2-3$ lower SFRs are found taking into account that in 
luminous quasars, like the {ones} observed at $z>6$, $30-60\%$ of the dust heating  
may be due to the AGN emission itself \citep{Wang10, Schneider15}. 
Indeed, \cite{Schneider15} show that the optically bright quasar J1148 may 
contribute $30-70\%$ of the observed FIR luminosity ($>20 \rm \mu m$) heating 
the large amount of dust ($\sim 3\times 10^8\msun$) in the host galaxy 
ISM. We refer the reader to \citet{V14} and \citet{Schneider15} for a 
discussion.}.

\subsection*{Is there a stellar mass crisis?}
The fast enrichment in metals and dust at very high redshift discussed above
suggests that quasar host galaxies {could} have undergone intense episodes of star 
formation. Similar chemical abundances {are typically found in} local galaxies which, 
however, evolved on longer time scales. 

{The estimated mean BH-stellar bulge mass ratio, 
$\rm M_{BH}/M_{star}$, of $z\sim 6$ quasars is about 10 times higher than the one 
observed in the local Universe (e.g. \citeauthor{Wang10} \citeyear{Wang10, 
Wang13}),
suggesting that high redshift BHs may have formed or assembled earlier than their 
host galaxies (e.g. \citealt{Lamastra10}).}
Although this result could be strongly affected by observational selection 
effects \citep{Lauer07, VolonteriStark11} and large uncertainties in the 
estimation of the mass and size of the stellar bulge (\citealt{V14, P16}), 
it is difficult to explain how the ISM has been enriched to chemical 
abundances similar to that of local galaxies, {albeit with $\lesssim 10\%$ of the 
stars \citep{V11, Calura14, V14}.}

\subsection*{What is the role of BH feedback?}

It is expected that galaxy-scale winds, triggered by the large amount 
of energy released in the BH accretion process, play a crucial role in 
regulating the BH-host galaxy co-evolution, shaping the SFH 
and BH accretion history itself
(e.g. \citealt*{SilkRees98}; \citealt{Granato04, DiMatteo05, Springel05} 
\citeauthor{Ciotti09} \citeyear{Ciotti09, Ciotti10}; 
\citealt*{HopkinsElvis10, ZubovasKing12}).

Indeed, massive and fast large scale gas outflows, associated to quasar 
activity, have been observed in local and high redshift quasars 
(\citeauthor{Feruglio10} 
\citeyear{Feruglio10, Feruglio15}; \citealt{Alatalo11, Aalto12, Alexander10}
\citeauthor{Nesvadba10}, \citeyear{Nesvadba10, Nesvadba11}, \citealt{Maiolino12, 
CanoDiaz12, Farrah12, Trichas12, Carniani16}).
At $z>6$ a massive gas outflow has been inferred from observations of [CII] 
emission line in J1148, revealing an outflow rate $\geq 2000-3000$ 
M$_\odot/$yr \citep{Maiolino12, Cicone15}.

{However there are still open issues like: what is the outflow powering mechanism, what are the effects of BH 
feedback on the host galaxy, how can the observed strong outflows and starbursts be simultaneously sustained? 
Although there are hints of star formation being quenched by quasar feedback 
at high redshift \citep{CanoDiaz12, Farrah12, Trichas12, 
Carniani16}, it is unclear if such feedback is able to completely suppress 
star formation in galaxies \citep{Peng15}.
On the other hand, it has been pointed out that AGN-driven positive feedback 
\citep{Zinn13, Cresci15} which triggers or enhances star formation, may be as 
important as quenching mechanisms in galaxy formation (e.g. \citealt{Gaibler12, 
Wagner13, Silk13, Bieri15}). }

\section{THEORETICAL MODELS}
\label{sec:models}
%

In the following sections we review the results of state-of-the art theoretical models for the formation of the first BHs, the properties of the environment in which they form and the evolution of their host galaxies. We focus on models in which the evolution of the baryonic component of galaxies is followed by means of analytic prescriptions linked to their host dark matter (DM) halo properties. 
In particular, we discuss two complementary approaches adopted to describe DM halos, 

\begin{itemize}
\item{\textit {pure semi-analytic models (pSAMs)}}: that use analytic algorithms (e.g. Monte Carlo) usually based on the extended Press-Schechter (EPS, \citep*{PS74, LC93}) or similar, formalism (see e.g. \citealt*{Parkinson08, SomervilleKolatt99, Zhang08})\\
\item{\textit {hybrid semi-analytic models (hSAMs)}}: that use cosmological N--body simulations (e.g. \citealt{Springel05Nature}) to extract DM halo properties (e.g. mass and spatial distribution) and build their models on top of them.
%
\end{itemize}

Pure semi-analytic techniques are commonly adopted to shed light either on the 
early gas enrichment with metals and dust in the high redshift ISM
(\citealt*{HF02, ME03, DGJ07}; \citealt{V09}; \citeauthor{Gall11a}  
\citeyear{Gall11a, Gall11b}, \citealt{DC11}, \citealt{Mattsson11, Pipino11, 
Calura14}) or on the origin of the first SMBHs and the resulting BH-host galaxy 
scaling relations (e.g. \citeauthor{Volonteri03} \citeyear{Volonteri03, 
Volonteri05}; \citealt{Madau04}; \citealt*{VolonteriRees06}, 
\citealt{Dijkstra08}; \citealt*{TH09}; 
\citeauthor{Devecchi10} \citeyear{Devecchi10, Devecchi12}; 
\citealt{Petri12, Dijkstra14}; \citealt*{VSD15}).

However, in order to interpret the observed properties of high redshift quasars discussed in the previous section it is important to connect all the physical processes regulating the formation of SMBHs and the host galaxies' chemical evolution history in a self-consistent cosmological framework.

A first attempt to link the chemical evolution of the ISM (metals and dust) to the SMBH formation in $z>6$ quasar by means of a pSAM has been made by \citeauthor{V11} (\citeyear{V11, V14, V16}) and \citet{P16} employing the cosmological data-constrained model \textsc{GAMETE/QSOdust}. The model successfully {reproduces the observed properties of a sample of $z>5$ quasars such as the mass of molecular gas, metals, dust and BHs \citep{V14} and has been recently improved to investigate different SMBHs formation scenarios. The relative role of light and heavy seeds is investigated in \cite{V16}, while \citet{P16} study the effect of different gas accretion modes/regimes by including new, physically motivated, prescriptions for gas cooling, disk and bulge formation in progenitor galaxies.}

These models are targeted to highly biased regions of the Universe, where a SMBH is expected to form (e.g. \citealt{Stiavelli2005,Kim2009,Utsumi2010,Morselli14}), namely single DM halos of $10^{12}-10^{13}\msun$, 
which represent the highest density fluctuations at $z\sim 6$ (e.g. \citealt{Fan04}; 
\citealt*{VolonteriRees06}). 
In other words, all the halos in the merger trees of high-z pSAMs are the ancestors of a single quasar host. In particular, the observed/inferred properties of the best (observationally) studied quasar, J1148 at $z=6.4$, are often adopted as a reference data set to constrain/explore model parameters (e.g. \citealt{DGJ07, V09, V11, DC11, V16, P16}) in the above mentioned studies.}

The importance of several physical processes has emerged from both pSAMs and hSAMs, such as metal enrichment of the medium from galactic winds \citep{Dijkstra14, Habouzit16hSAM} and the clustering radiation sources \citep{Dijkstra08,Agarwal12}. The dependence of these physical aspects on the spatial halo distribution is {better described by} hSAMs as cosmological simulations: either DM only or hydrodynamical, directly provide the spatial distribution of halos.
In general, hSAMs {are designed to describe average volumes of the Universe that are able to probe smaller scales, exploring in detail the environmental conditions required for the formation of the high redshift BH population}.

The population of SDSS quasars presents an observational limit of 1 cGpc$^{-3}$ for $10^9\msun$ BHs (e.g. \citealt{Fan06, Venemans13, deRosa14}). Much larger volumes, and thus large scale N--body simulations 
are required to produce one such billion solar mass BH in a statistically significant manner, from either a Pop~III or a DCBH seed. 
On the other hand, small scale N--body simulations (i.e. much smaller volumes 
$\sim 100\ \rm cMpc^{-3}$)
are instead best suited for studying the environment in which the first stars and seed BHs form.
Either way, hSAMs operating on either of these volumes present complementary insights into the problem of forming BHs at $z>6$. 

%

So far, hSAMs have mostly been used to study the formation of heavy seeds. For example, \citet{Agarwal12, Habouzit16hSAM} use hSAMs in which DM only {simulations permit one to account for effects that are critical to the first galaxy formation paradigm. Local feedback mechanisms such as the net radiation flux and metal pollution can be folded into the construct of hSAMs, along with other recipes such as self--consistent star formation and tracking halo histories across cosmic time.}

The first part of this review is dedicated to the description of the environmental conditions required for the formation of different populations of seed BHs in both average volumes, simulated by hSAMs, and biased regions {described in pSAMs. We then will briefly discuss different pathways for the fast growth of these seeds up to
$>10^9\msun$  BHs at $z\sim 6$, as well as their co-evolution with the host galaxies.}

\section{The first seed BHs: how, where and when}
\label{sec:seedBHs}
In the following sections we discuss the environmental {conditions that enable and regulate} the formation of the first seed BHs in a cosmological context, as explored by both pSAMs and hSAMs. We focus our attention on the formation of light (Pop~III remnants) and heavy (DCBHs) seeds.

\subsection{Seeds formation sites}
\label{sec:seedsSites}
As they are the end products of massive Pop III stars, light {seed} 
formation is enabled by nearly primordial conditions: metal and dust poor gas fragmenting into one or few massive stars at redshift $z\sim 20$ (e.g. \citealt{Abel02, Heger03}; \citealt*{MadauRees01}; \citealt{Yoshida08, Latif13b, Hirano14}). Gas enriched up to metallicity {$Z_{\rm cr}\geq10^{-4}$\zsun}, or dust-to-gas ratios $\cal{D}$ $>4\times 10^{-9}$, fragments more efficiently (thanks to metal lines cooling and dust continuum radiation), to form instead lower mass, population II (Pop II) stars (\citealt{Schneider02, Schneider03, Omukai05, Schneider12}).
Such conditions are expected to be easily met in the first virialised 
structures at early times, the so-called minihalos, characterize by virial temperatures of $1.2\times 10^3<T_{\rm vir}<10^4$ K  and masses $M_{\rm h}\sim 10^{5-6}\msun$ (see e.g. \citealt{Bromm13} for a review) 

Pop~III stars also represent the first sources of light and heavy elements (including dust, e.g. \citealt{Nozawa07, HegerWoosley10, Marassi15}), setting the stage for all subsequent structure formation in their neighborhood. Therefore, it is imperative that their formation is captured in the models for a consistent identification of the seed BH hosts. 
Resolving minihalos, in which these stars form, is thus crucial for models, at least at $z>20$. Unfortunately, the mass/size resolution limit in both hSAMs (i.e. the box size and DM particle mass) and the pSAMs (i.e. the minimum DM halo mass) is often determined by the inherent computational costs.


Depending on the aim of the model, different scale/mass resolutions are suited for different studies. 
Resolving arbitrarily small halos is computationally prohibitive even for analytic binary Monte Carlo algorithms. In pSAMs the resolution of the merger tree is thus defined by the minimum halo mass, which, together with the adaptive redshift interval ($\Delta z$) are chosen to maintain manageable computational times, simultaneously matching the EPS predictions at different redshifts (e.g. \citealt{Volonteri03}; \citealt*{TH09}). 

In N--body simulations, the need to resolve a minihalo sets an upper limit on the box-size that can be simulated in a reasonable time frame.
N--body simulations with volumes $\sim 100 \rm cMpc^3$ allow {one to} resolve minihalos, 
capturing the small-scale sub-grid physics. These simulations offer insights on the formation sites of the first stars and seed BHs but {lack statistical significance in terms of SMBH adundance for which larger volumes are required as discussed in section \ref{sec:models}.}
\newline

\noindent The formation of a DCBH requires the absence of 
star formation and of efficient coolants (metals and dust) in order to maintain isothermal collapse of gas clouds in Lyman$-\alpha$- cooling halos (Ly$\alpha$, $T_{\rm vir}\sim 10^4 \ \rm K$), leading to a Jeans halo mass (which scales as $T^{3/2}$) which is high enough to avoid fragmentation. 
Thus, heavy seed BHs are expected to form out of poorly enriched gas ($Z<Z_{\rm cr}$) if star formation is inhibited by an intense H$_2$ photo-dissociating flux, i.e. photons in the Lyman Werner (LW) band ($11.2-13.6$ eV) emitted by {nearby external sources. These} conditions indeed enable the formation of a supermassive star (SMS) that may eventually lead to a massive seed BH by accreting the surrounding material (e.g. \citealt{BL03, Begelman06}; \citealt*{SpaansSilk06}; \citealt*{IO12}; \citealt{Inayoshi14, Ferrara14}).

This peculiarity of the environmental conditions, and the frequency of their occurrence is still under debate (\citeauthor{Agarwal12} \citeyear{Agarwal12, Agarwal14}; \citealt{Habouzit16hSAM, Dijkstra14, Chon16}) .
The conditions are sensitive to galaxies' assembly histories and on the interplay between the effect of chemical, radiative and mechanical feedback, driven by star formation and BH growth itself.


\subsection{Forming the first stars}
\label{sec:stars}

{In star forming halos both Pop~III or Pop~II stars form depending on the chemical enrichment (metallicity) of the gas. Pop~III stars form out of metal-free/poor gas ($Z<Z_{\rm cr}$) while metal/dust-rich gas clouds instead lead to Pop~II star formation. }

The metallicity of a galaxy is usually the result of the interplay between in-situ and external metal {pollution, i.e. stellar nucleosynthetic products injected in the galaxy interstellar medium (ISM), and {in--falling} metal rich (and dusty) gas ejected from nearby galaxies via supernovae (SNe) and AGN-driven winds.}

\begin{figure}
\includegraphics[width=0.8\columnwidth]{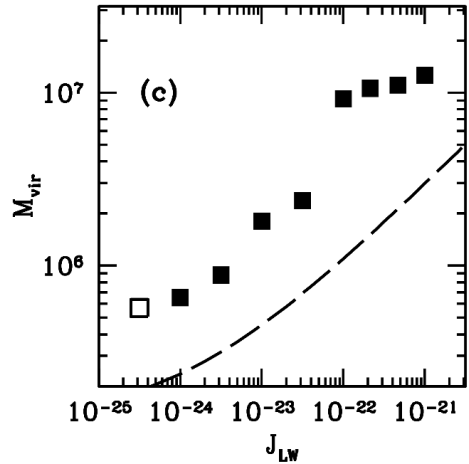}
\caption{The $M_{\rm crit}-J_{\rm LW}$ relation from \citet{OsheaNorman08}, Fig. 3. The squares represent their updated calculations while the \citealt{Mach01} relation is depicted by the dashed line. If the mass of a pristine minihalo exposed to a given $J_{\rm{LW}}$, lies above the curve formed by the squares, it is considered Pop III star forming.}
\label{fig:popIIImh}
\end{figure}

Most hSAMs allow Pop~III stars to form in metal--free halos, i.e. the ones that have never hosted a star in their past and/or pass the critical mass threshold \citep{Agarwal12}.
{The mass threshold can be understood as a negative feedback effect of LW photons that delay Pop III SF by a fraction of dissociating H$_{\rm{2}}$ molecules in a minihalo. While exposed to LW radiation}, $J_{\rm LW}$\footnote{Note that we use the term flux and specific intensity interchangeably in the manuscript where both refer to a specific intensity in the LW band in units of $10^{-21}\rm erg^{-1}s^{-1}cm^{-2}Hz^{-1}sr^{-1}$}, the halo must grow (or accrete more gas) in order to replenish the H$_{\rm{2}}$ content, thereby becoming suitable for Pop III SF (e.g. \citealt{Mach01, OsheaNorman08}). We show this $M_{\rm{crit}}-J_{\rm{LW}}$ curve expressed as Eq.~\ref{eq:mJvirial} (\citealt{Agarwal12}), in Fig.~\ref{fig:popIIImh} (from \citealt{OsheaNorman08}), where

\begin{eqnarray}
M_{\rm crit} \approx 4 \left(1.25 \times 10^5 + 8.7 \times 10^5 \left({ 4 \pi J_{\rm LW} } \right)^{0.47}\right). \ 
\label{eq:mJvirial}
\end{eqnarray}


In their recent pSAMs, \citet{V16} and \citet{deBennassuti17} compute the fraction of gas that can cool down and form stars in minihalos as a function of halo virial temperature, redshift, gas metallicity and level of LW flux $J_{\rm LW}$ at which the halo is exposed. 
At a given redshift, the halo mass threshold increases with $J_{\rm LW}$. Progressively more massive minihalos are expected to form stars at lower redshifts, at a fixed $J_{\rm LW}$. A value $J_{\rm LW}\leq 0.1$ is already high enough to suppress star formation in the less massive minihalos ($<(3-4)\times 10^6\msun$) at $z>20$. 
In good agreement with the gas collapse simulations of \citet{OsheaNorman08}, Pop~III star formation is inhibited in $\leq 10^7 \rm M_\odot$ pristine ($Z=0$) minihalos exposed to a LW flux $J_{\rm LW}\geq 1$, at redshift $z<17$. Stronger $J_{\rm LW}$ levels (e.g. $>10$) sterilize all pristine minihalos already at redshift $z=20.$\footnote{Note that \citet{V16} and \citet{deBennassuti17} also investigate the dependence of the $M_{\rm crit}-J_{\rm LW}$ relation on metallicity. They show that the presence of a small amout of metals does not significantly affect the results as long as $Z\leq 10^{-1.5} \rm Z_\odot$. At higher metallicities, gas cooling and thus star formation can occur in progressively smaller halos so that $\sim 10^6 \rm M_\odot$ minihalos are able to form stars already at $z\lesssim 20$ (we refer the readers to the original papers for more details).}

To date, observations do not provide strong enough constraints on the Pop~III IMF. On the other hand, theoretical studies provide predictions on the mass distribution of these stars, that varies among different study (see e.g. the reviews by \citealt{Bromm13, Glover13}). 

The most commonly adopted scenario in hSAMs (e.g. \citealt{Agarwal12, Agarwal13,Chon16}) is to form 1 Pop III star in a minihalo, randomly picked from a top--heavy IMF that ranges from $100-1000\msun$. For atomic cooling (Ly$\alpha$) pristine halos that are suitable for Pop III star formation, generally a cluster of $10-100$ stars are allowed to form, following the same IMF. 

{Regardless of the DM halo mass, massive Pop~III stars with an average mass of $\sim 100-200\msun$ are allowed to form in high-z pSAMs} (e.g. \citealt{V11, V14, P16}). The number of stars depends on the total stellar mass formed in each star formation episode, and thus on the star formation efficiency and available gas mass. An alternative scenario for Pop~III star formation in pSAMs has been proposed by \citet{V16}: Pop~III stars form with an intrinsic top-heavy IMF in the mass range $[10-300] \, \rm M_\odot$. Then, this IMF is stochastically sampled, on the fly, according to the time-dependent total mass of newly formed stars.
We will discuss the effect of these two different assumptions for Pop~III stars formation on the light seed BHs distribution, later (in Fig.~\ref{fig:seedBHs}).

In metal-rich halos, Pop~II star formation is generally accounted for by converting a fixed fraction of the available gas into stars. 
The time/redshift evolution of the gas content is modeled either by scaling the DM halo mass with the universal baryon fraction (e.g. \citealt{Dijkstra08,Dijkstra14,Habouzit16hSAM}) or solving a set of differential equations (e.g. \citealt{V11,V14,V16, Agarwal12, P16}). In hSAMs the star formation recipes are usually calibrated to reproduce the cosmic star formation rate density (CSFRD) observed at $z>6$ \citep{Hopkins04,Mannucci07,Bouwens08,Laporte12}. 
Since pSAMs are generally targeted to explain the existence of a single quasar, the models are designed to match the observables of the quasar in question.

\subsection{Conditions for direct collapse}
\label{sec:nDCBHs}
The treatment of the DC {scenario} is now taking advantage of hybrid models where instead of Press-Schechter merger trees, one uses a fully cosmological N--body simulation as a playground for the various recipes critical to DCBH formation.
One of the main advantages of using hSAMs to study the formation of SMBHs at early times is the spatial information {that enables one to study the dependence of various processes on the halos' physical distribution within the simulated volume.}
Nearby star-forming halos emit LW photons that are able to photo-dissociate $\rm{H_{2}}$ \citep{Omukai01,Omukai08,Shang10,Latif13a}, and thus the spatial distance between halos is a crucial ingredient as it controls the strength of the irradiation flux (e.g. \citealt{Agarwal16a})

We provide here an overview of the large scale feasibility of the DC model, i.e. we do not consider studies related to the formation of individual DCBHs (see e.g. \citealt{LatifFerrara16} for a review), and rather discuss studies which aim at deriving statistical properties, such as the number density of DCBH sites that form in the early Universe and the conditions leading to them.

In order to identify a DCBH formation site within an average volume of the Universe, one must account for the entire LW and metal pollution history of the atomic cooling halo in question, especially taking into consideration the effects of the local environment. This is one of the biggest strengths of hSAMs as painting galaxies on N--body simulations allows us to compute spatial locations.

\subsubsection{Critical LW flux}
We have discussed above how (low level) LW flux can delay Pop III star formation in pristine minihaloes. Once the halo becomes atomic cooling, i.e. when it attains a virital temperature of $T_{\rm{vir}} > 10^4\ \rm K$ and the primary coolant becomes atomic hydrogen \citep{Omukai2000}, an extremely high level of flux can completely shut down H$_{\rm{2}}$ cooling by dissociating these molecules in the most dense (thus efficiently self--shielded) regions \citep{Omukai01,Omukai08,Shang10,Latif13a}.

The critical level $J_{\rm cr}$, above which direct collapse of gas clouds into massive seeds is enabled, is still a matter of debate and remains a free parameter for models.
Assuming that Pop III stellar populations mimic a $T = 10^5\, \rm K$ and Pop II stellar populations a $T = 10^4\, \rm K$ blackbody, \citet{Omukai2000} computed the critical value of $J_{\rm{crit}}$ using their 1D spherically symmetric gas collapse model. Since the shape of the blackbody spectrum depends on its temperature, $J_{\rm{crit}}$ depends on the type of the stellar population externally irradiating the pristine atomic cooling halo.
They found $J_{\rm{cr}}^{\rm{III}} \approx 10^4 - 10^5$ and $J_{\rm{cr}}^{\rm{II}}\approx 10^2-10^3$ is needed from Pop III and and Pop II populations to cause DCBH formation in a neighboring pristine atomic cooling halo. Revisions in this estimate followed by employing high resolution 3D hydrodynamical simulations and better recipes for H$_{\rm 2}$ self--shielding, leading to an estimate of $J_{\rm{cr}}^{\rm{III}} \sim 1000$ and $J_{\rm{crit}}^{\rm{II}}\approx 10-100$ \citep{Shang10,WolcottGreen2011,Latif14b, Hartwig15}. 
 
In addition, ionizing photons and X-rays can both increase the free electron fraction promoting $\rm H_2$ formation (\citealt*{IO11}; \citealt{Yue14, Johnson14}; \citealt{IT15}). As a result a higher critical LW level, up to $J_{\rm cr}\sim 10^4-10^5$, is required (\citealt{Latif14b, Regan14}; \citealt*{LatifVolonteri15}).

Besides H$_{\rm{2}}$ molecules, H$^{\rm -}$ ions play a critical role in pristine gas collapse as they regulate H$_2$ formation at densities $n\lesssim 10^{3} \ \rm cm^{-3}$ via the reactions

\begin{eqnarray}
&\rm H^{ }& +\ {\rm e} \rightarrow \rm H^- + \gamma \label{reac.hmfromh} \\
&\rm H^-& +\ \rm H \rightarrow \rm H_2 + e^{-} \label{reac.h2fromhm}
\end{eqnarray}

The importance of this network is further understood by their corresponding photo--destruction channels

\begin{eqnarray}
&\rm H_2& +\ \gamma_{\rm LW} \rightarrow \rm H + H \label{reac.pdi} \\
&\rm H^-& +\ \gamma_{0.76} \rightarrow \rm H + e^{-} \label{reac.pde}
\end{eqnarray}
where $\gamma_{\rm LW}$ and $\gamma_{\rm{0.76}}$ represent the photons in the LW band and {photons with energy greater than} $0.76$ eV respectively. Ignoring the role of 1eV photons can lead to a gross overestimation in the value of LW flux required to suppress H$_2$ cooling, as demonstrated by \citet{WolcottGreen2011} and \citet{Haiman2012}. Furthermore, \citet{Glover15a, Glover15b} showed that inconsistencies in the chemical networks and reaction rate coefficients can lead to a factor $\sim 3$ difference in the determination of $J_{\rm{cr}}$. 

The assumption of representing Pop III and Pop II spectral energy distributions (SEDs) as blackbodies was questioned by \citet{Sugimura14,Agarwal2015, Agarwal16a} who showed that using realistic SEDs to represent stellar populations instead drastically alters the paradigm. This is because the change in the slope of a SED with the age of a stellar population alters the rate of production of LW photons with respect to 1eV photons. \citet{Agarwal16a,Wolcott2017} demonstrated that indeed, one can not expect a single value of $J_{\rm{cr}}$ from a given stellar population, but that it is a value dependent on the underlying stellar population's SFH and varies from $0.1 - 1000$ in their 1D models. Needless to say, given that these studies are very recent, this variation in the nature of $J_{\rm{cr}}$ needs to be further explored. 

\begin{figure*}
\begin{center}
{\includegraphics[width=0.8\textwidth]{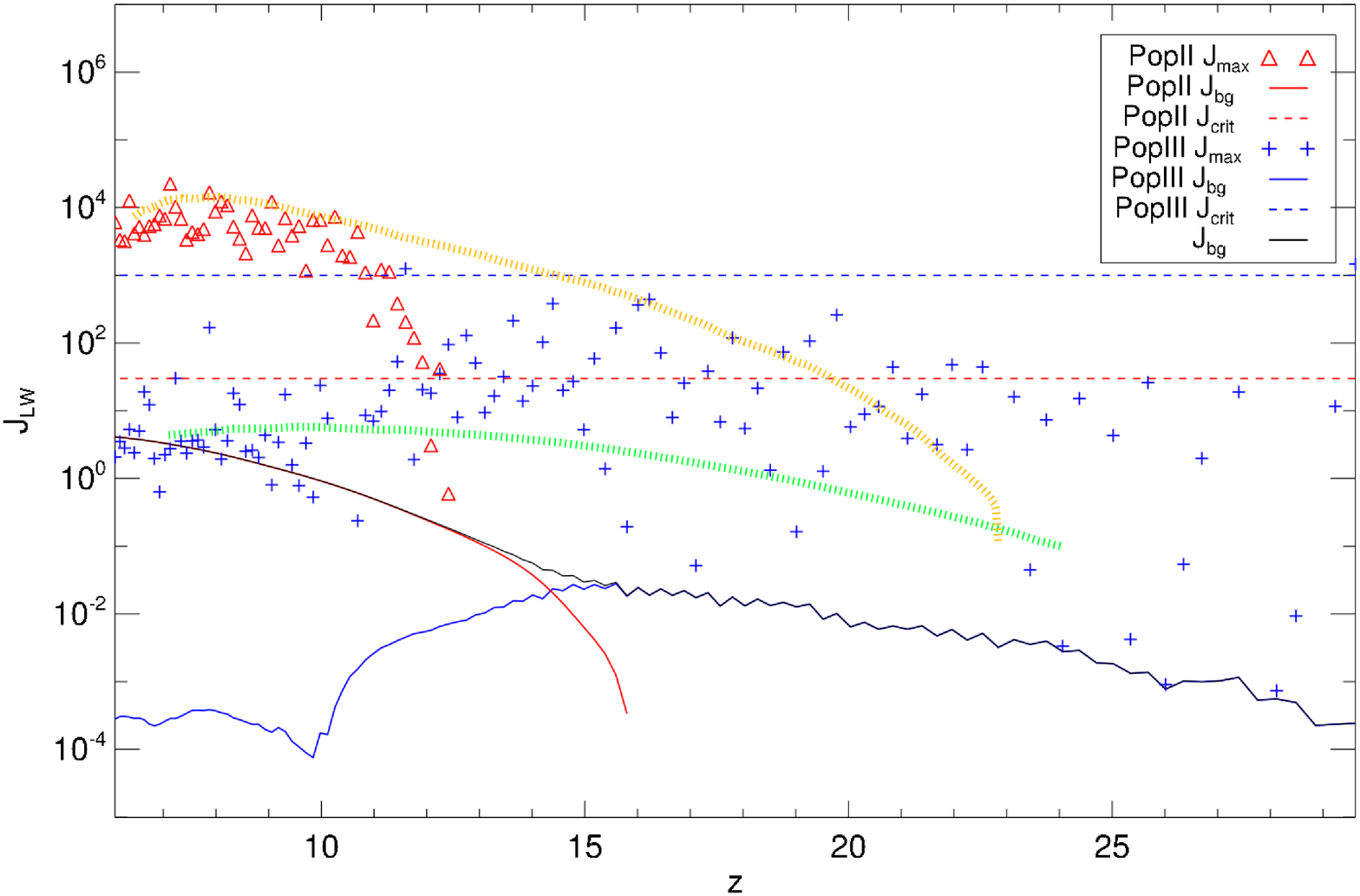}}
\caption{From \citet{Agarwal12}: The background and local level of LW radiation plotted for each redshift. ``The red triangles ($J_{\rm local}^{\rm II}$) and blue crosses ($J_{\rm local}^{\rm III}$) indicate the maximum value of LW radiation to which a pristine halo is exposed at each redshift in their volume. The red and blue dashed lines represent $ J^{\rm II}_{\rm crit}$ and $J^{\rm III}_{\rm crit}$ respectively. It is interesting to see that the maximum value of $J_{\rm local}^{\rm III}$ (blue crosses) falls short of $J^{\rm III}_{\rm crit}$ (blue dashed line). However, in the case of Pop II sources, the maximum value of $J_{\rm local}^{\rm II}$ (red triangles) is several orders of magnitude higher than the $J^{\rm II}_{\rm crit}$ (red dashed line)." 
Finally, the orange dotted line shows the average LW emission from \citet{V16}.}
\label{fig:jcomp}
\end{center}
\end{figure*}

In Fig.~\ref{fig:jcomp} we show the global and spatial LW intensities from the \citet{Agarwal12} hybrid fiducial model, and compare them to other studies. The averaged background LW intensity,$J_{\rm bg}$, at a given redshift is computed as a function of the stellar mass density at that redshift.
\begin{equation}
J_{\rm bg}(z) = \frac{hc}{4\pi m_{\rm H}}\eta_{LW}\rho_\star(1+z)^3\ , \nonumber
\end{equation}
where $\eta_{\rm{LW}}$ is the number of LW photons emitted per stellar baryon, and $\rho_\star$ is the stellar mass density at a given redshift, $z$. Both quantities are linked to the stellar population, so that $J_{\rm bg}=J^{\rm III}_{\rm bg}+J^{\rm II}_{bg}$ (see \citealt{GreifBromm06, Agarwal12} for more details).
The green dotted line is instead the specific intensity $J_{\rm bg}$ given by \citet{Dijkstra14}. 
The orange dotted line in Fig.~\ref{fig:jcomp} shows the average LW emission computed in the pSAM of \citep{V16} (similar values are also shown by \citet{Petri12}).

As it can be seen from the figure, the global LW background radiation, $J_{\rm bg}$ is always far below the critical value for DC, $J_{\rm cr}$ (horizontal dashed {red and blue} lines). 
Thus, the study of the spatial variation of the photo-dissociating emission is fundamental to identify potential DCBH formation sites. 

\citet{Ahn09} presented the first study of the evolution of the inhomogeneous LW background, in which the local LW flux intensity is self--consistently computed in a cosmological N--body simulation, explaining its importance. Their study is based on a suite of runs that were originally aimed at understanding reionization \citep{Iliev07}, but was modified to include a radiative--transfer module for LW photons. \citet{Ahn09} find that the average intensity of the LW radiation exceedes the threshold value for $\rm H_2$-cooling and star formation suppression in minihalos well before the reionization process is complete. In their scenario, both the average and local LW flux can be $\geq 10^{-2}$ already at $z<20$ (see e.g. their figure 10). As a result, $Ly\alpha$-cooling halos are the dominant sources of reionzation while minihalos are sterilized before they can significantly contribute to the ionizing and LW background radiation.
Following this study, 
several other models (pSAMs and hSAMs) pointed out the importance of LW flux fluctuations due to sources clustering in the formation of DCBHs (e.g. \citealt{Dijkstra08, Dijkstra14, Agarwal12, Habouzit16hSAM, Chon16, Pawlik14}). 

In Fig.~\ref{fig:jcomp} we also show the values of the local LW flux, $J_{\rm{local}}$, from single stellar populations as computed by \citet{Agarwal12} in their hSAM volume at each redshift. They show that while Pop~III stars are never able to produce the $J_{\rm{crit}}^{\rm{III}}$ in their vicinity, Pop II stars are able to produce $J_{\rm{crit}}^{\rm{II}}$ quite easily (see \citealt{Agarwal12} for details). This result was later  confirmed by \citet{Agarwal14, Habouzit16hydro} in their suite of hydrodynamical runs, and by \citet{Chon16}.

Due to the lack of spatial information, pSAMs instead can not capture the spatial variations of $J_{\rm LW}$ with respect to the background flux as hSAMs do. 
However, the LW emission from Pop~III/II stars and accreting BHs is self-consistently computed, according to their SED, as a function of stellar age and metallicity and of BH accretion rate (e.g. \citealt{Petri12, V16}). An important difference with respect to hSAMs is that in pSAMs the star formation and BH accretion efficiency are usually calibrated to match the observed SFR and BH mass of specific, single, obejcts (e.g. quasar J1148 in \citealt{V16}). 
Within the biased region occupied by the progenitors of a $10^{13}$ DM halo, the computed LW flux can be interpreted as a mean value for the local fluctuations exceeding the background level, as expected by several models (e.g. \citealt{Dijkstra08, TH09, Agarwal12, Dijkstra14}). 
In addition, \citet{Petri12} and \citet{V16} show that stellar emission provides the dominant contribution to the photo-dissociating flux with respect to accreting BHs. For example, the global LW emission from stellar populations in \citet{V16}, taken as a proxy of the local flux in their biased region (orange dotted line Fig.~\ref{fig:jcomp}), is in good agreement with the maximum local Pop~II LW flux, at $z<11$ (red triangles), and with the large scatter in the maximum local Pop~III flux, at larger redshifts (blue crosses), from \citet{Agarwal12}. 



\subsubsection{The role of metal enrichment}

As the first generation of stars form in the Universe they also create the first wave of metals that provide the conditions for Pop~II star formation. Thus, it is critical to understand metal pollution in terms of both in--situ and external effects. The chemical enrichment of a given halo is indeed the result of the ongoing and past star formation (i.e. metals and dust produced by stars in the parent galaxy and/or its progenitors) as well as contamination by infalling material from outside the halo (galactic winds). Both self-enrichment and winds play a role in setting the conditions for {seed BH} formation. 

As we have seen above, several models (both pSAMs and hSAMs) point out that DCBH regions are expected to be close to star-forming galaxies, in order to maintain a low abundance of $\rm{H_{2}}$. These are also the first regions which are exposed to metal-pollution from galactic winds driven by SNe (and AGN).

Although \citet{Agarwal12} do not explicitly consider galactic winds in their model, their results on the number and environment of DCBH sites were in good agreement with the FiBY suite of hydrodynamical simulations \citep{Agarwal14} that did include external metal pollution. This suggests that, for the assumed $J_{\rm cr}=30$, the DCBH population is not significantly affected by winds. {\bf Using their analytic approach \citet{AgarwalCR7_2} find that even with instantaneous metal mixing, the metal outflows (e.g. due to SN winds) in the irradiating galaxy are unable to prevent the advent of isothermal collapse in the neighbouring DCBH halo. The external atomic cooling site has sufficient time to undergo isothermal collapse in the presence of the LW radiation field before being polluted by the metal outflow.}


\citet{Dijkstra14} explore the effect of metal pollution by both SN-driven galactic outflows and genetic enrichment on the DCBH formation {probability }
by computing the size of regions that can be enriched with metals transported by galactic SN-driven winds and the probability that a halo remains metal free (i.e. it do not inherit metals from its progenitor halos). They show that external metal pollution sterilizes DCBH host candidates on a scale of $\leq 10$ kpc. The results suffer from the lack of spatial information in their pSAM.

The effect of galactic winds has been recently confirmed by \citet{Habouzit16hSAM}.
In their model, DC is enabled in the vicinity of $\sim 10^{11}$ M$_\odot$ star-forming halos, that can provide a high enough radiation intensity ($J_{\rm LW}>J_{\rm cr}=100$, see their Fig.~3) to halos at a distance of $\sim 15-20$ kpc at $z>15$, without polluting them. In other words when the expanding metal rich bubbles created by SN explosions are still smaller than the regions irradiated by a strong intensity.\\
By means of a set of differential equations \citealt{V11, V14, V16} self-consistently follow the global life cycle of the mass of metals and dust in the ISM of J1148 progenitor galaxies taking into account the metal pollution (infalls) of the external medium due to both SN- and AGN-driven winds. They find that a more efficient self-enrichment of galaxies within a merger tree, with the respect to the average genetic pollution history, may prevent the formation of DCBHs progenitors before the LW flux exceeds the critical threshold, while infalling metals are responsible for the super-critical enrichment of newly virialised halos (see e.g. \citealt{V16}).

It is worth noting that, metal mixing is an extremely complicated topic. The time scale for metals escaping their host halo and mixing with the gas of the halo being polluted is not fully understood (e.g. \citet{CenRiquelme08, WiseAbel08, Smith15}). Additionally, the escape of metals from their parent halo depends on the wind--escape--velocity and the potential well of the halo \citep{Muratov2015,Smith15}.

Metal-enrichment is indeed predicted to be very disparate in the early Universe, but some halos could remain metal-free down to $z\sim 6$ \citep{Tornatore07, Pallottini14}. The fraction of metal-free halos, or at least halos below the critical metallicity to avoid fragmentation, depends on chemical and mechanical processes \citep{Schneider06a,Schneider06b}.
Detailed prescriptions of the effects of inhomogeneous enrichment  as well as of the physical properties of metal winds escaping from star--forming halos can not be easily modeled in either pure or hybrid SAMs. However cosmological hydrodynamical simulations can self-consistently track the evolution of metal-enrichment over the entire simulated volumes \citep{Latif16,Agarwal14,Habouzit16hydro}. 

Summarizing, the combined effect of chemical and radiative feedback sets the condition for the formation of both light and heavy seeds as it regulates Pop~III/II star formation in all halos and determines the fraction of atomic cooling (Ly$\alpha$) halos that can potentially host DCBHs at later times.
As long as the build up of a super-critical $J_{\rm LW}$  precedes the efficient metal pollution, DCBH formation can occur in atomic cooling halos.

\subsection{DCBHs number density}

Over the past few years, the question of the number density of DCBHs has become a topic of great interest, and has led to values that span several orders of magnitude, from $\sim 10^{-1}$ to $10^{-9}$ cMpc$^{-3}$.

Here we compare the results of both hSAMs \citet{Agarwal12, Habouzit16hSAM} and pSAMs of \citet{Dijkstra08, Dijkstra14, V16}. 
We include DCBH number densities from the \citet{Agarwal14} and \citet{Habouzit16hydro} hydrodynamycal simulations as they offer a direct comparison of semi-analytic and hydrodynamic approaches.

\citet{Dijkstra08} compute the probability distribution function of the LW flux at which DM halos are exposed to at $z\sim 10$ taking into account their clustering properties. They find that only a small fraction, $<10^{-6}$, of all atomic cooling halos are exposed to a LW flux exceeding the assumed critical threshold, $J_{\rm LW}>10^3$ and thus derive a number density of $<10^{-6}$ cMpc$^{-3}$ potential DCBHs hosts.

In contrast, using a semi-analytic model on top of a cosmological N--body simulation, \citet{Agarwal12} find a higher number density, in the range $10^{-2}-10^{-1}$ cMpc$^{-3}$ for $\rm{J_{crit}}=30$ (their fiducial model), even accounting for in--situ metal pollution from previous star formation events.

In their fiducial model, \citet{Dijkstra14} include star formation in atomic cooling halos (but do not include minihalos), metal pollution from progenitor halos and galactic outflows and estimate $n_{\rm DCBH}\sim 10^{-9}-10^{-6}$ cMpc$^{-3}$ between $z=20$ and 7. They explore the dependence of their predictions on model assumptions, such as the value of LW photons escape fraction and critical flux for DC, underlying the important effect of galactic winds decreasing the number density by several orders of magnitudes.

More recently, \citet{Habouzit16hSAM} find consistent results, with a number density of DCBH regions in the range $10^{-7}-5\times 10^{-6}$ cMpc$^{-3}$. A factor of 2 higher number density can be found in cosmological N--body simulations in which primordial fluctuations are described by a non-Gaussian distribution. In addition they also estimate the Pop~III remnant BHs number density, being about 2 order of magnitude higher than that of DCBHs, although they do not resolve minihalos in their simulations.
Similar values are found in hydrodynamical simulations by \citet{Habouzit16hydro} for
different box sizes and resolutions.

In their pSAM aimed to study the role of Pop~III remnant BHs and DCBHs in the formation of a $z\sim 6$ SMBH, \citet{V16} predict an average number density of $\sim 10^{-7}$ cMpc$^{-3}$ DCBHs. These are the DCBHs expected to form in J1148 progenitor galaxies, along the hierarchical history of a $10^{13}$ DM halo. As we will discuss later, only a fraction of these heavy seeds eventually end in the final SMBH, driving its fast growth.

\begin{figure*}
\begin{center}
   \includegraphics[scale=0.8]{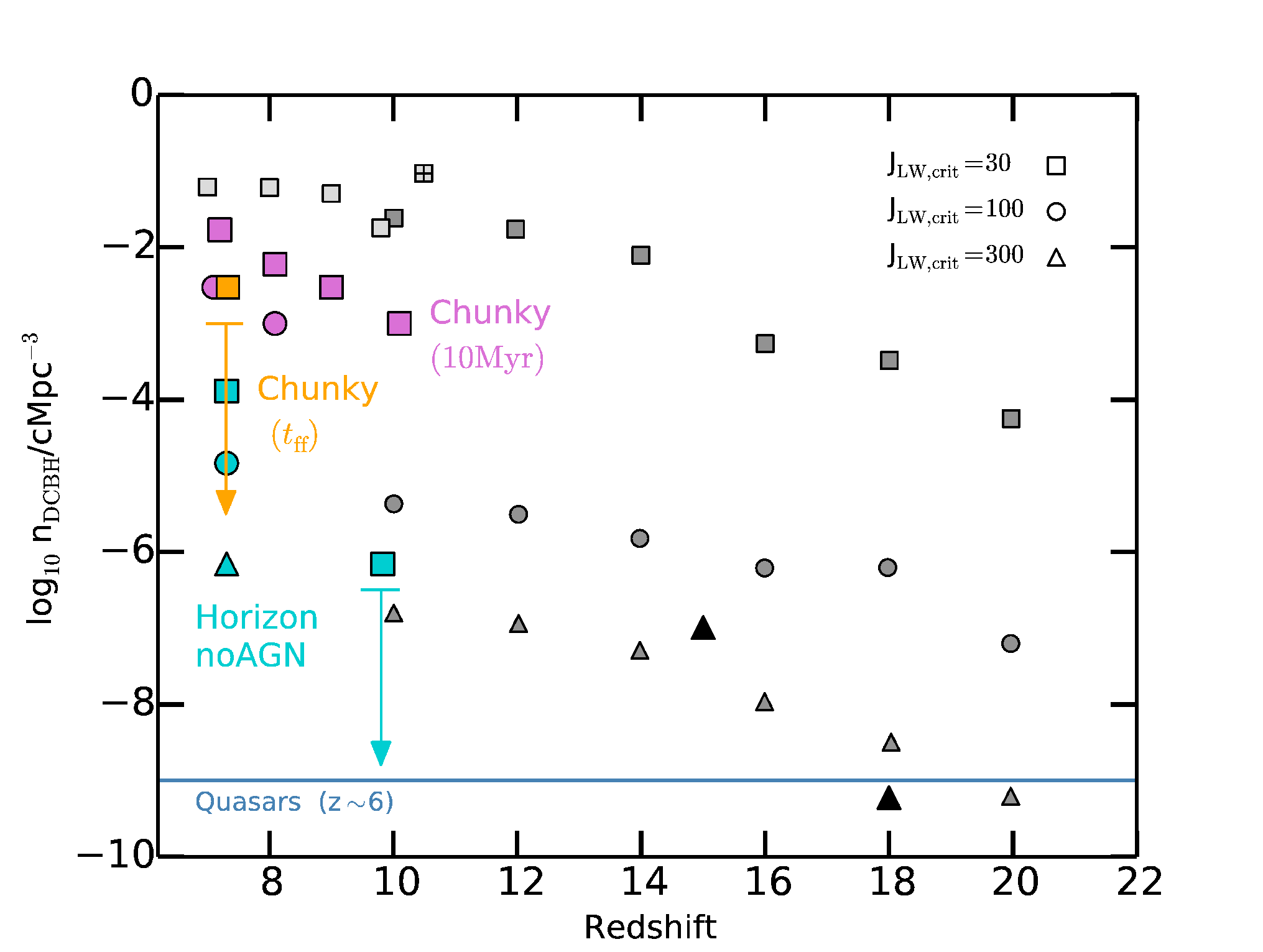}
\caption{Comoving number density of halos that can host a DCBH, at a given redshift. Symbols represent different radiation intensity thresholds. Squares:  $\rm{J_{LW,crit}}=30$, circles:  $\rm{J_{LW,crit}}=100$, triangles:  $\rm{J_{LW,crit}}=300$. The horizontal solid blue line show the comoving number density of $z\sim 6$ SMBHs. The light gray crossed square at $z=10.5$ is from the hydrodynamical simulation by \citet{Agarwal14}, the light gray squares in the range $z=10-7$ are from \citet{Agarwal12} (private communication), dark gray squares and black triangles are the results of \citet{Dijkstra14} and \citet{V16}, respectively. The orange square shows the number density for \citet{Habouzit16hydro} (10 cMpc side box, $\rm{t_{ff}}$, see text). The purple squares and circles show the number density for \citet{Habouzit16hydro} (10 cMpc side box, 10 Myr, see text). The cyan squares, circle and triangle represent the large-scale cosmological simulation Horizon-noAGN \citep[][142 cMpc side box]{Dubois14c, Habouzit16hydro}.}
\label{fig:number_density}
\end{center}
\end{figure*}

In Fig.~\ref{fig:number_density} we show a collection of DCBH number densities derived from some of the studies discussed above. Symbols represent different radiation intensity thresholds: squares refer to $\rm{J_{LW,crit}}=30$, circles to $\rm{J_{LW,crit}}=100$, and triangles to $\rm{J_{LW,crit}}=300$. The figure is taken from \citet{Habouzit16hydro} who compare the results of semi-analytic studies by \citet{Dijkstra14} (dark gray symbols) with hydrodynamical simulations: 
one of the the FiBy simulations based on the smoothed particle hydrodynamics (SPH) code \textsc{gadget} (e.g. \citealt{Springel05}) presented by \citet{Agarwal14} (light grey crossed square at $z=10.5$);
two runs of the 10 cMpc box Chunky simulation with a collapse times scale equal to 10 Myr (purple symbols) and to the halo free fall time, $t_{\rm ff}$ (orange square); 
the large-scale (142 cMpc side box) cosmological simulation Horizon-noAGN (cyan symbols, \citealt{Dubois14c,Peirani17}). We refer the reader to the original paper \citet{Habouzit16hydro} for a detailed discussion. We have included in this figure the predictions by \citet{Agarwal12} in the $z=7-10$ redshift range (light gray squares) and \citet{V16} at $z=18$ and 15 (black triangles). 
Finally, the horizontal blue solid line show the SMBH number density observed at $z\sim 6$ of 1 cGpc$^{-3}$.

\subsubsection{Consensus between different studies}
One of the most restrictive ingredient of the DC scenario is the absence of $\rm{H_{2}}$ (through both $\rm{H_{2}}$ destruction and prevention of $\rm{H_{2}}$ formation){ to keep} the gas temperature and thus the Jeans mass high enough to avoid the fragmentation of gas clouds. This should favor the formation of only one massive object. The presence of strong LW radiation is then required to strongly depress $\rm{H_{2}}$ abundances \citep{Omukai2000,Omukai08,Shang10}.
From \citet{Ahn09}, we have understood that the spatial variations of the radiation intensity, driven by LW photons able to photo-dissociate $\rm{H_{2}}$, was certainly a key requirement of the scenario. Most of the models for the radiation intensity include now a spatial varying component based on local photo-dissociating sources. The radiation intensity is either computed directly from stellar particles according to their age, distance, and redshift \citep{Agarwal12, Agarwal14, Habouzit16hydro}, or from the stellar mass painted on DM halos \citep{Dijkstra14,Habouzit16hSAM, Chon16}. \\

\begin{figure}
\centering
   \includegraphics[width=8cm]{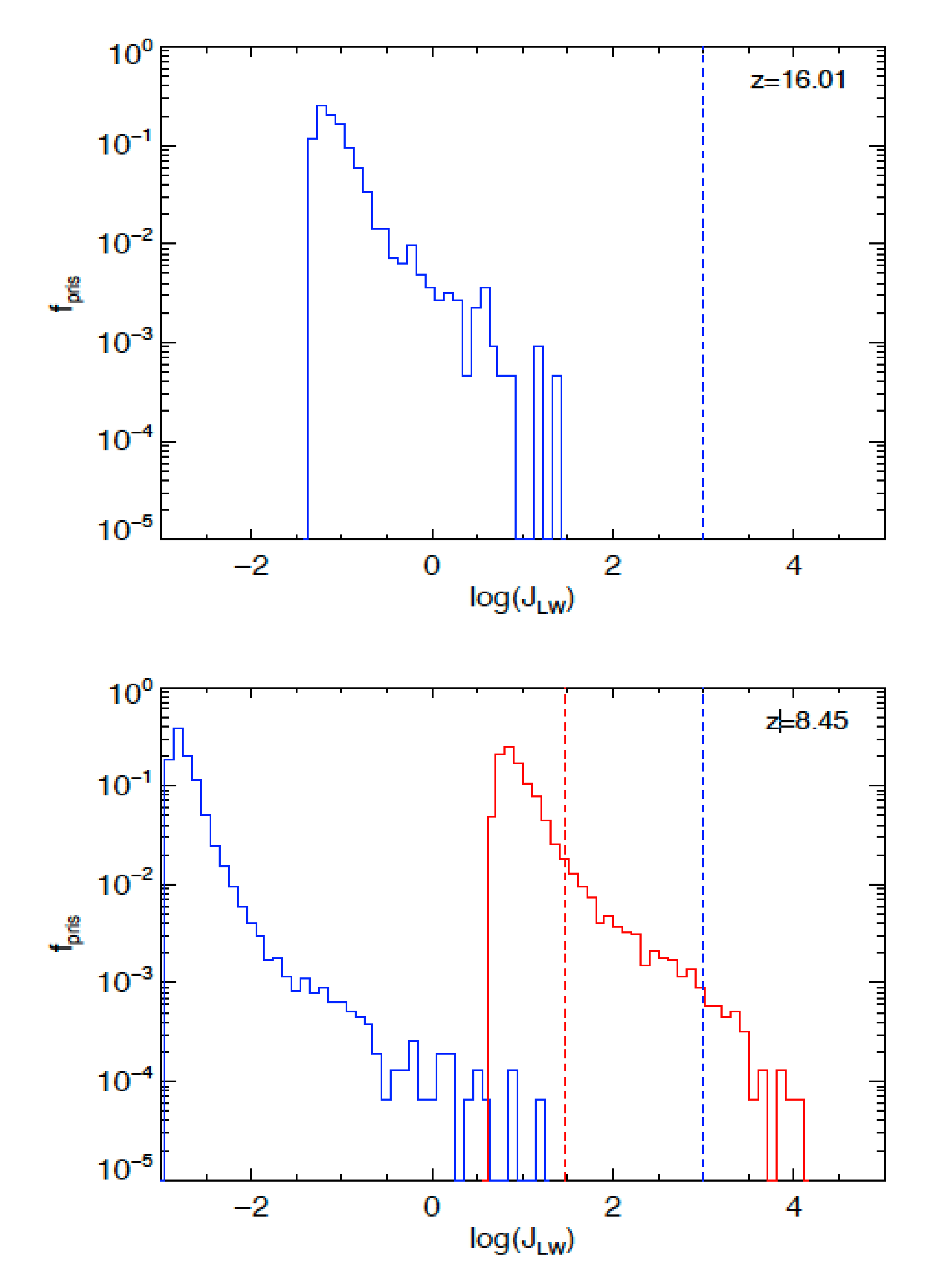}
\caption{Distribution of local radiation intensity \citep{Agarwal12} seen by pristine halos at $z=16$ (top panel), before the formation of Pop~II begins, and later on at $z\sim9$ (bottom panel) when Pop~II is already in place. Radiation intensity from Pop~III stars is shown in red, and radiation intensity from Pop~II stars in blue. Dashed lines indicate the critical radiation intensity expected for Pop~III stars (in red) and Pop~II stars (in blue). Pop~III stars radiation intensity appears to be almost always below the critical intensity (below the corresponding dashed line), whereas a fraction of pristine halo illuminated by Pop~II stars radiation flux can meet the critical radiation intensity condition.}
 \label{fig:pdf_jlw}
\end{figure}

\noindent Moreover, the critical radiation flux needed to destroy $\rm{H_{2}}$, seems to be driven mainly by Pop~II stars. This is supported by three main ideas. First of all, the LW radiation background created by Pop~III stars emission, impacts their surrounding by photo-dissociating molecular hydrogen. Cooling rate decreases, which delays the gas collapse, and this vicious circle lowers and delays the formation of new Pop~III stars at later time \citep{OsheaNorman08, Johnson12}.
The life time of Pop~III stars is also thought to be short ($\sim10$ Myr), it could be too short for providing a high LW radiation intensity during the whole free-fall time of a dark matter halo. One can compute the redshift at which the free-fall time is approximately equal to $\sim10$ Myr, and finds $z\sim45$. This means that a halo illuminated only by Pop~III radiation, could form a BH only at very early times, around $z\sim 45$. Finally, the intensity of Pop~III radiation itself may be not enough to provide the critical radiation intensity commonly assumed for the DC model \citep{OsheaNorman08,Agarwal12}. 
In Fig.~\ref{fig:pdf_jlw} \citep[reproduced from ][]{Agarwal12}, we show the distribution of the local varying radiation intensity seen by pristine halos at $z=16$, before the formation of the first Pop~II stars, and $z\sim 9$, after their formation. {Radiation intensity from Pop~III stars is shown in blue, and from Pop~II stars in red. Dashed lines indicate the critical radiation intensity expected for Pop~III stars (in blue) and Pop~II stars (in red). Pop~III stars radiation intensity appears to be almost always below the critical intensity (below the corresponding red dashed line), whereas a majority of pristine halos under Pop~II stars radiation flux can meet the critical radiation intensity condition.}
The distribution of radiation intensity to which halos are exposed to, is in good agreement between various studies, using similar methods and LW radiation modelings \citep{Agarwal12,Chon16}, or different approaches \citep{Dijkstra08}. 

Finally, all studies agree that metal-pollution from both heritages, previous episodes of star formation in halo progenitors and galactic winds from nearby halos, play a fundamental role. Galactic winds are able to sterilize potential DCBH regions by enriching them in metals, on a scale of $\leqslant 10\,\rm{kpc}$, reducing by one order of magnitude the number density of DCBHs \citep{Dijkstra14}. However, candidate halos for the DC model are close to LW sources, within the same characteristic length of $\sim10\, \rm{kpc}$ \citet{Agarwal12}.
{There is now consensus on the fact that halo clustering favours the emergence of DCBH regions, as it leads to a radiation field higher than the average background \citep{Dijkstra08,Agarwal12,Agarwal14}.}

\subsubsection{Why do we have a spread in the number density}
{The large diversity of models (modeling of the photo-dissociating radiation intensity, and metal-enrichment, for example), methods (pSAMS, hybrid with DM only simulations, or hydrodynamical ones), set-up of simulations (star formation, SN feedback), used to estimate the number density of DCBH regions, complicate the task of comparing their results.} Despite the fact that all the studies presented here seem to agree pretty well, several of the models use different assumptions. In this section, we identify the main {differences} between the different models.\\
\citet{Habouzit16hydro} perform a comparison between the SAM model of \citet{Dijkstra14} and the hybrid model of \citet{Agarwal14}, and find that compared to hydrodynamical simulations, \citet{Dijkstra14} overestimates the stellar mass that form in halos. In the opposite, \citet{Dijkstra14} underestimate the number of galaxies that contribute to radiation, and the extent of metal polluted bubbles (the latter can vary strongly depending on the stellar mass going SN, and the medium properties). {In some cases, the different assumptions compensate each other, and lead to the same estimate of the number density of the potential DCBH host halos \citep{Habouzit16hydro}.}\\

Differences between models using dark matter only simulations and models from hydrodynamical simulations can be studied by comparing \citet{Agarwal12} (distribution of halos from a dark matter simulation) and \citet{Agarwal14} (hydrodynamical simulation). The number density derived by \citet{Agarwal12} is shown in light gray squares in Fig. \ref{fig:number_density}, whereas the number density from \citet{Agarwal14} is represented in crossed square point in Fig. \ref{fig:number_density}. \citet{Agarwal14} is an improvement of \citet{Agarwal12}, because now, thanks to the hydrodynamical output, the model takes into account self-consistently cooling of halos, metal-enrichment through SN feedback, molecular dissociation and photo-ionization. 

As discussed above hSAMs are largely adopted to study the feasibility of the DCBH formation scenario. However, one would eventually want to know whether these heavy seed BHs, that formed at early times, can actually grow and form the population of quasar we see at $z=6$, and under which conditions this is possible (accretion, galaxy-galaxy mergers, super-Eddington episode, and so on).\\

Most of the studies discussed in this review provide upper limits on the number density of DCBHs, because they are not able to follow all the physical processes from the selection of dark matter halos to the collapse of the gas to form a BH. However, they seem to all show that the DCBH number density is higher than the observed number density of quasars at high redshift, $10^{-9} \,\rm{cMpc^{-3}}$, horizontal blue line in Fig.~\ref{fig:number_density} (\citealt{Fan06, Mortlock11}). {If a higher critcal flux is required for DCBH formation ($\rm{J_{crit} > 100}$), as it is actually found in 3D zoom--in simulations, then \citet{Dijkstra14} \citep[see also][with the large scale simulation Horizon-noAGN]{Habouzit16hydro} show that the upper limit on the DCBH number density is sufficient to reproduce the population of quasars. However, such high critical values do not explain the population of less massive BHs that we observe today in more normal and low-mass galaxies \citep{Greene12, Reines13}. }

{On the other hand, smaller simulation boxes that resolve minihalos and include a more developed chemistry network, have lead to the derivation of higher DCBH region number density, particularly because they impose a lower critical radiation intensity ($\rm{J_{crit}}=30$) \citep{Agarwal12,Agarwal14}.} Such low values of the critical intensity could suggest that the DC scenario may also be able to seed the more {{\textit{normal}}} galaxies. Recently, \citet{Habouzit16hydro} show that this also strongly depends on SN feedback implementation, and that to explain BHs in normal galaxies, a weak SN feedback is required.

Although large progress has been made, both in terms of pure SAMs and hybrid models to investigate the DC scenario, owing to the the large spread in the number density of DCBH regions derived, and the uncertainty in the nature of the critical LW radiation intensity, it is still unclear if the DC scenario can produce enough BHs to explain the population of high redshift quasars. 

Regarding the target of this review, high redshift quasars, a natural follow up of these studies would be to follow the growth of the BHs, modeling the accretion and feedback as a function of host halo merger history. To this aim, a number of semi-analytic studies have been developed so far (see section \ref{sec:models}). In the following part of the manuscript we will review state-of-the-art results on the growth of $z\sim 6$ SMBHs and their host galaxies.

\section{From seeds to the first quasars}
\label{sec:quasars}

\begin{figure*}
\begin{center}
\includegraphics[width=6.cm]{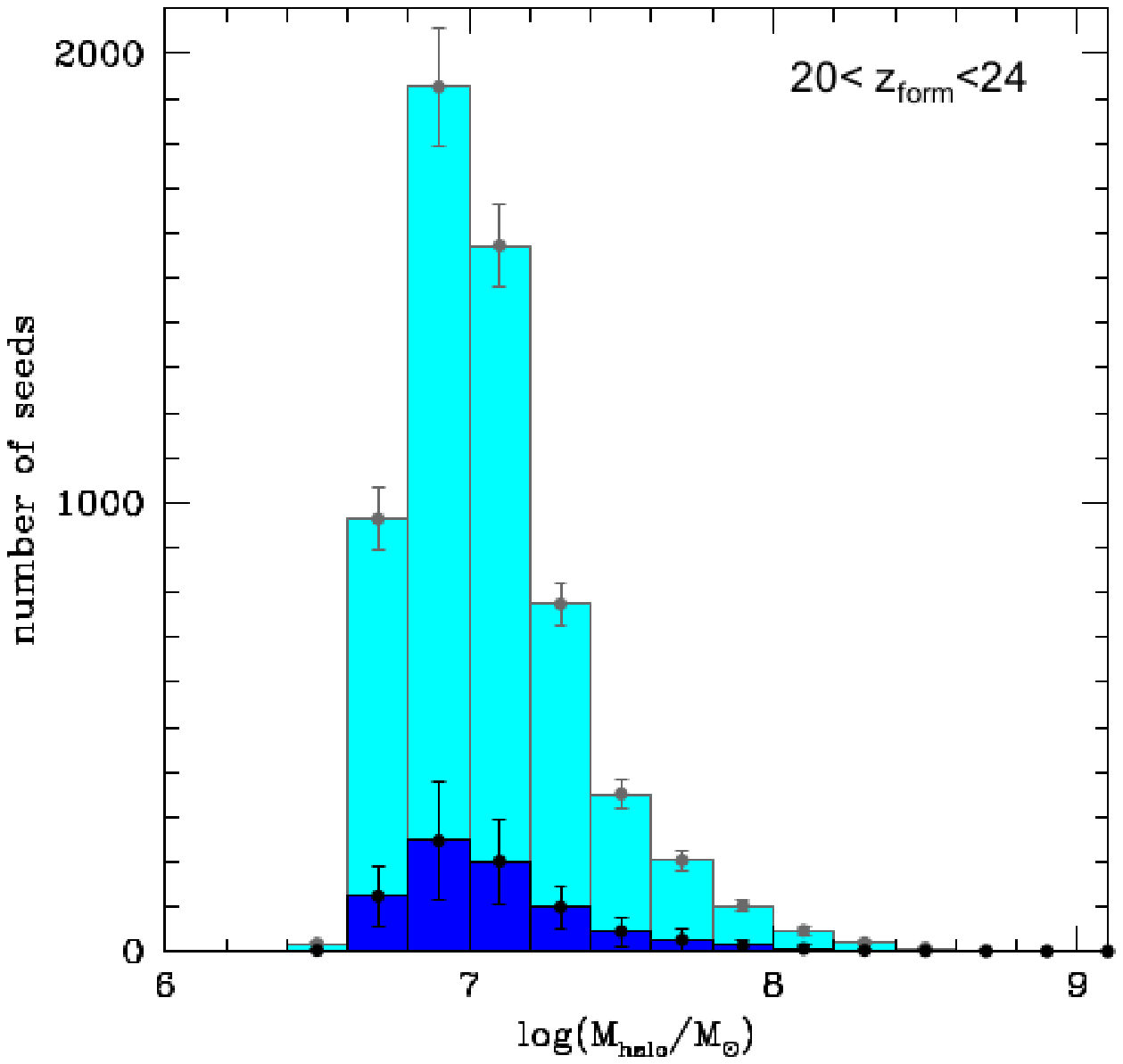}
\includegraphics[width=5.5cm]{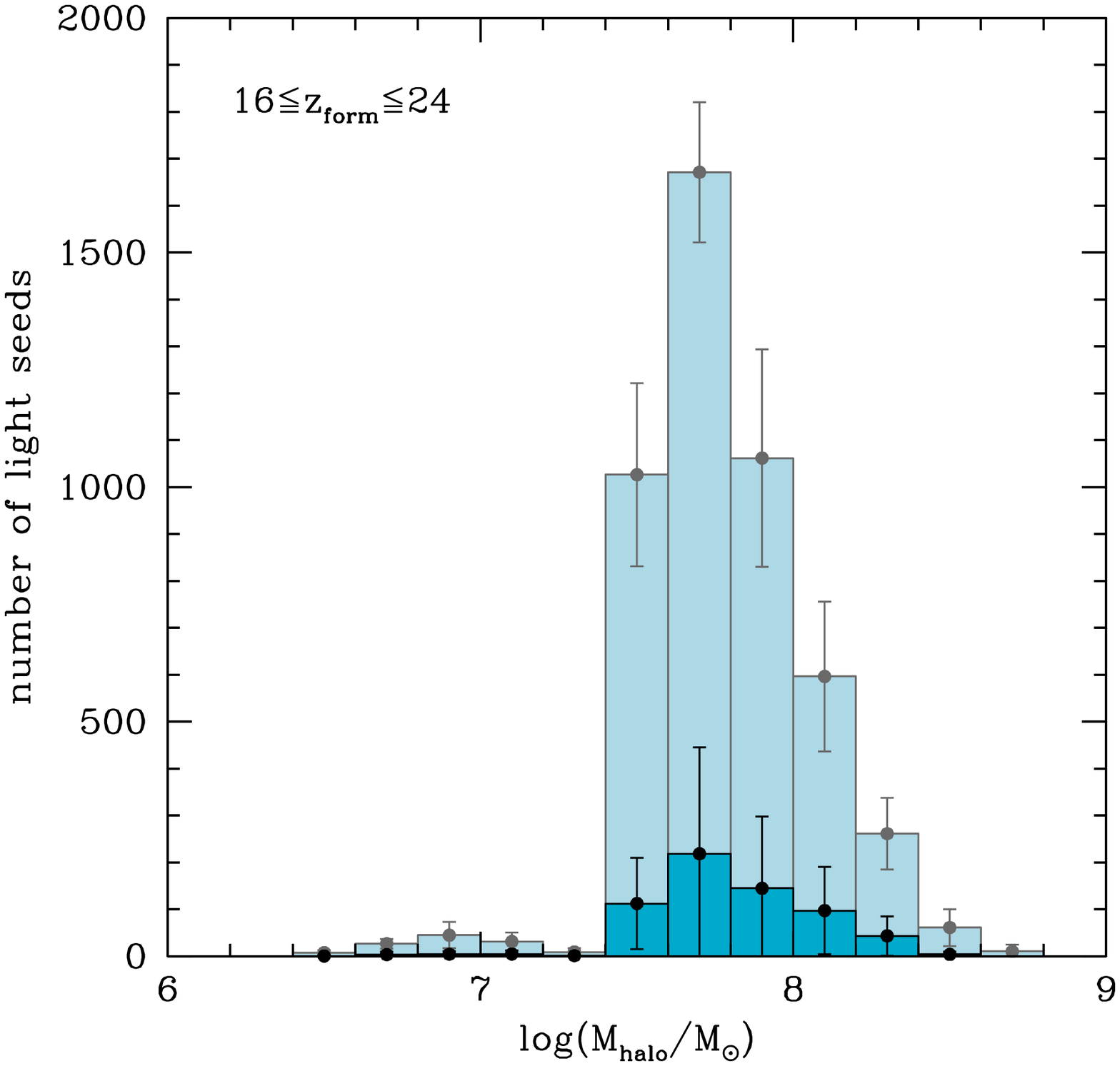}
\includegraphics[width=5.5cm]{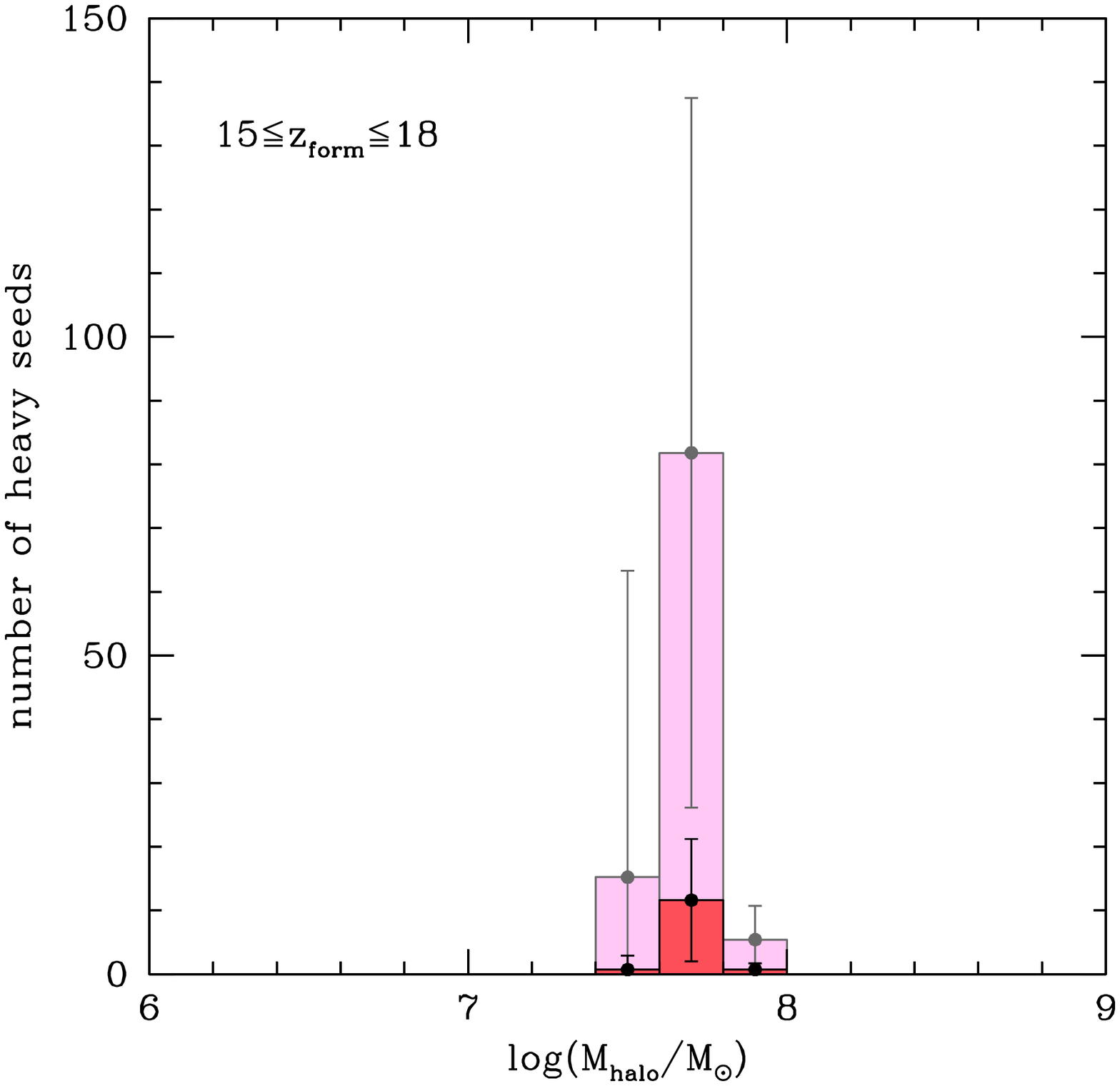}
\caption{Distribution of the average number of seed BHs as function of the 
DM halo mass from different seeding prescriptions adopted in pSAMs: 
\textit{(i)} equal-mass $100\msun$ light seeds (left panel) and
\textit{(ii)} (10-140) and (260-300)$\msun$ Pop~III remnant BHs (middle 
panel) plus $10^5\msun$ heavy seeds (right panel). Histograms and data 
points show the number of total (in lighter colors) and real SMBH progenitors 
(darker histograms, see text). Error bars account for the $1\sigma$ dispersion. 
The figures are adapted from \citep{P16} and \citep{V16}. The average 
redshift range in which seeds form, acording these two models, is given in each panel.
}\label{fig:seedBHs}
\end{center}
\end{figure*}
 
Several studies have investigated the early growth of SMBHs starting from either 
light or heavy seeds (see reviews by e.g. \citealt{Volonteri10, Natarajan11};
\citealt*{VolonteriBellovary12, AH12, Haiman13, JohnsonHaardt16}). 
In these models, SMBHs growth is driven by both gas accretion and mergers with other BHs.
In this section we briefly review the most recent studies of the hierarchical assembly of a quasar and its host galaxy, as described by pSAMs.

\subsection{Light vs heavy seeds}
\label{sec:LvsH}
{In Fig.~\ref{fig:seedBHs} we show} the distribution of the number of seed BHs formed along the hierarchical build up of a $z\sim 6$ quasar (i.e. in its progenitor galaxies) as a function of the host DM halo mass. In the left panel we show the number of equal mass stellar BHs, light seeds of $10^2 \msun$, assumed to form in newly virialized halos, as long as they are metal poor, $Z<Z_{\rm cr}=10^{-3.8}$, i.e. at $z\geq 20$, as predicted by \citet{P16}. The other two panels instead are for a mixed-seed-based seeding prescription \citep{V16}: $(40-140)$ and $(260-300)\msun$ Pop~III remnant BHs (middle panel) plus $10^5\msun$ heavy seeds (right panel), forming along the same merger history. In this scenario the formation of light and heavy seeds is simultaneously explored thus, allowing to directly compare the role of the two channels in the formation of a SMBH.
In all panels, histograms and data points are obtained by averaging over 29(10) different merger histories of the $10^{13}\msun$ DM halo in the light-seed(mixed-seed) case, with error bars showing the $1\sigma$ dispersion. Both prescriptions have been adopted to model the quasar, J1148 at $z=6.4$, with a SMBH of $(2-6)\times 10^9\msun$ (\citealt{Barth03, Willott03, deRosa11}). 
As we discussed in the previous sections, the number, redshift and typical host halo mass of both light and heavy seeds is determined by the interplay between the early chemical enrichment -- due to metal-rich infalling gas from the external medium, polluted by SN- and AGN-driven winds from other galaxies -- and the intensity of the LW radiation (from both stars and accreting BHs) {to which the halos are exposed}.

The inclusion of radiative feedback effects results in a less efficient and slightly slower metal enrichment, enabling Pop~III stars to form on average down to lower redshift, e.g. $z\sim 16$ in the model shown on Fig. ~\ref{fig:seedBHs}. { As we see in the right panel of the} figure, DCBH form in $10^7-10^8\msun$ progenitor halos (and in the narrow redshift range $15-18$, see \citealt{V16} for details), consistent with what is expected from their formation theory and the findings of \citet{Bellovary11, Agarwal12, Habouzit16a, Chon16}. 

In their pSAM, \citet{Petri12} combine both light and heavy seeds to investigate their relative role in the formation of SMBHs in a pSAM. They explore the dependence of the resulting SMBH evolutionary scenario on the fraction of halos (exposed to a LW flux $J_{\rm cr}>10^3$) that can actually host DCBHs. A $10^9-10^{10}\msun$ BH is formed at $z\sim 6$ if at least $(1-10)\%$ of all the halos host a heavy seed (see their Figs.~4 and 9).

For a critical LW threshold $J_{\rm cr}>300$ \citet{V16} predict an average heavy seeds occurrence ratio (the number of galaxies with $Z<Z_{\rm cr}$ when $J_{\rm LW}>J_{\rm cr}$ divided by the number of all the halos exposed to a flux $J_{\rm LW}>J_{\rm cr}$) of $\sim 5\%$ at $z>15$. This suggests that chemical feedback plays a dominant role in determining of the birth environment\footnote{Indeed, if for example, a factor of $\sim 4$ higher $J_{\rm cr}$ is assumed in this model, the formation of heavy seeds is completely suppressed by chemical feedback.}. 

Recently, \citet{Chon16} combined a semi-analytic model for galaxy formation with halo merger trees extracted from N-body DM simulations to select possible DCBH hosts among atomic cooling halos. By means of zoom--in cosmological hydrodynamical simulations of the {selected halos, they explore the evolution of gas collapse in the DCBH sites. They mostly follow the approach of \citet{Agarwal12} but bring a previously unexplored effect to light: tidal gravitational fields affecting gas collapse.} They show that unless assembled via major--mergers, their DCBH sites do not survive the tidal fields and get disrupted before an isothermal collapse can ensue at gas densities of $n\geq 10\ \rm cm^{-3}$. A DCBH occupation fraction of $\sim 5\%$ (2 out of the selected 42) is found in this study, in good agreement with the pSAM of \citet{V16}. 

\subsection{The role of mergers and BH dynamics}

{Merger events can serve as an important physical process that drives the growth of BHs.}
However, binary (or multiple) BH interactions, driven by dynamical friction, 
are quite complex, multi-scale processes.
The physical scales of interest {span} from sub-pc scales of the 
Schwarzschild radius (e.g. $\sim 10^{-11}$ pc for $100-300\msun$ BHs and 
$\sim 10^{-8}-10^{-7}$ pc for BHs of $10^5-10^6\msun$) up to the Mpc scale 
of the host galaxy mergers. 
In addition, the mechanism leading to BH-BH mergers, the time 
it takes for BHs to coalesce via gravitational wave (GW) emission, and the relation between the end--state of the merger and the properties of the respective host galaxies, are still open questions.

However, SAMs aimed to study the formation and evolution of SMBHs trough cosmic 
time usually adopt simple prescriptions to account for the contribution of mergers 
to the BH growth (see e.g. \citealt{TH09} and references therein). 
 
In major mergers\footnote{Usually major and minor mergers are defined according to the 
mass ratio of the two merging DM halos (e.g. \citealt*{TH09} and reference therein). 
For example, a mass ratio higher than $1:10$ is assumed by \citealt*{VolonteriRees06} to identify major merger events.
}
BHs follow the fate of their host galaxies, coalescing 
to form a more massive BH. 
However, during this process, a large center-of-mass recoil (kick)
can be imparted to the newly formed BH as a consequence of asymmetric gravitational 
wave emission (e.g. \citealt{Campanelli07, Schnittman08, Baker08}). 
The acquired kick velocity can be as large as $\sim 100 \rm\ km s^{-1}$,
enough to eject the coalesced binary out of the host galaxy (see e.g. \citealt{Yoo04}; 
\citealt*{VolonteriRees06}; \citealt*{TH09}; \citealt{Barausse12} and references therein for details).
On the other hand, in minor mergers one of the two merging BHs, usually the 
least massive one, is assumed to remain as a satellite in low-density regions, 
without accreting or contributing to the growth of the final BH.

The effective number of seed BHs from which a SMBH forms depends on these assumptions.   
\citet{V16} predict that only $\sim 13\%$ of the light and heavy seeds in their model (darker histograms in middle and right panels in \ref{fig:seedBHs}) contribute to the final mass of the SMBH of J1148, at $z=6.4$, as a large fraction of BHs is lost due to minor mergers.

A similar fraction, $\sim 15\%$ (indicated by the darker histogram in the left panel) is left by taking into account the combined effect of minor mergers and gravitational recoil on growing light seeds. On average, $\sim 56\%$ satellites BHs are lost along the entire merger tree in minor mergers while $\sim 32 \%$ of the coalescing BHs, in major merger events, gain a recoil velocity large enough to exceed the retention speed, being kicked out of the galaxies (\citealt{P16}; a much larger fraction, $\sim 99\%$ is found by \citealt{Volonteri03}).

The effect of BH recoil due to gravitational wave emission during BH mergers has been also studied by \citet{Sijacki09}. They resimulate the most massive $z=6$ DM halo extracted from the Millennium simulation in order to study the effect of BH mergers \citep{Blecha16} in the growth of high redshift massive BHs. A SMBH of $10^9-10^{10}\msun$ is produced in an Eddington-limited scenario, by planting massive BH seeds of $10^5\msun$, in DM halos with masses $10^{9-10}\msun$ at z=15. They find that if the initial BH spin is high the growth of mostly isolated (only a small number of mergers occur) massive BHs is hampered. However, BH kicks substantially expel low-mass BHs, and thus do not affect the overall growth of the SMBHs. 

BH mergers are found to play a minor role in the formation of the first SMBHs 
(at relatively lower redshifts), in pSAMs (e.g. Fig 6 in \citealt{P16})
and recently in hydrodynamical simulations like MassiveBlack and BlueTides 
(e.g. \citet{Feng14, DiMatteo16}).\\
Mergers between BHs drive the black hole mass assembly only at high redshifts
(but see \citealt{Petri12}).
For example, although driving the BH growth process at $z>11$, BH-BH coalescences contributes to less than $1\%$ of the J1148 final BH mass at $z=6.4$ \citep{V11}.
Similarly, in \citet{V16}, BH mergers (of mainly light seeds) are predicted to 
drive the BH growth down to $z\sim 15$, before the gas accretion regime triggered by the formation of the first heavy seeds, sets in. 

Conversely, in the large-volume, cosmological hydrodynamical simulation Horizon--AGN (box size of 100 $h^{-1}$ Mpc and resolution mass of $8\times 10^7\msun$) \citet{Dubois14} 
show an accretion-dominated BH growth at high redshift, while in the older Universe, the galactic centers tend to be less gas-rich, and, thus, the mass growth of the central BHs is mostly driven by mergers.
In addition, a demographic study of BHs has been recently carried out by \citet{Volonteri16} within the same simulation. They show that the fraction of BH host galaxies is higher at higher stellar masses and that multiple BHs are hosted in the most massive halos as a consequence of merger events. A population of dual AGN, a central and an off--center accreting BH, is found in the simulated halos.

\subsection{The role of gas accretion}

\begin{figure}
\begin{center}
\includegraphics[width=8cm]{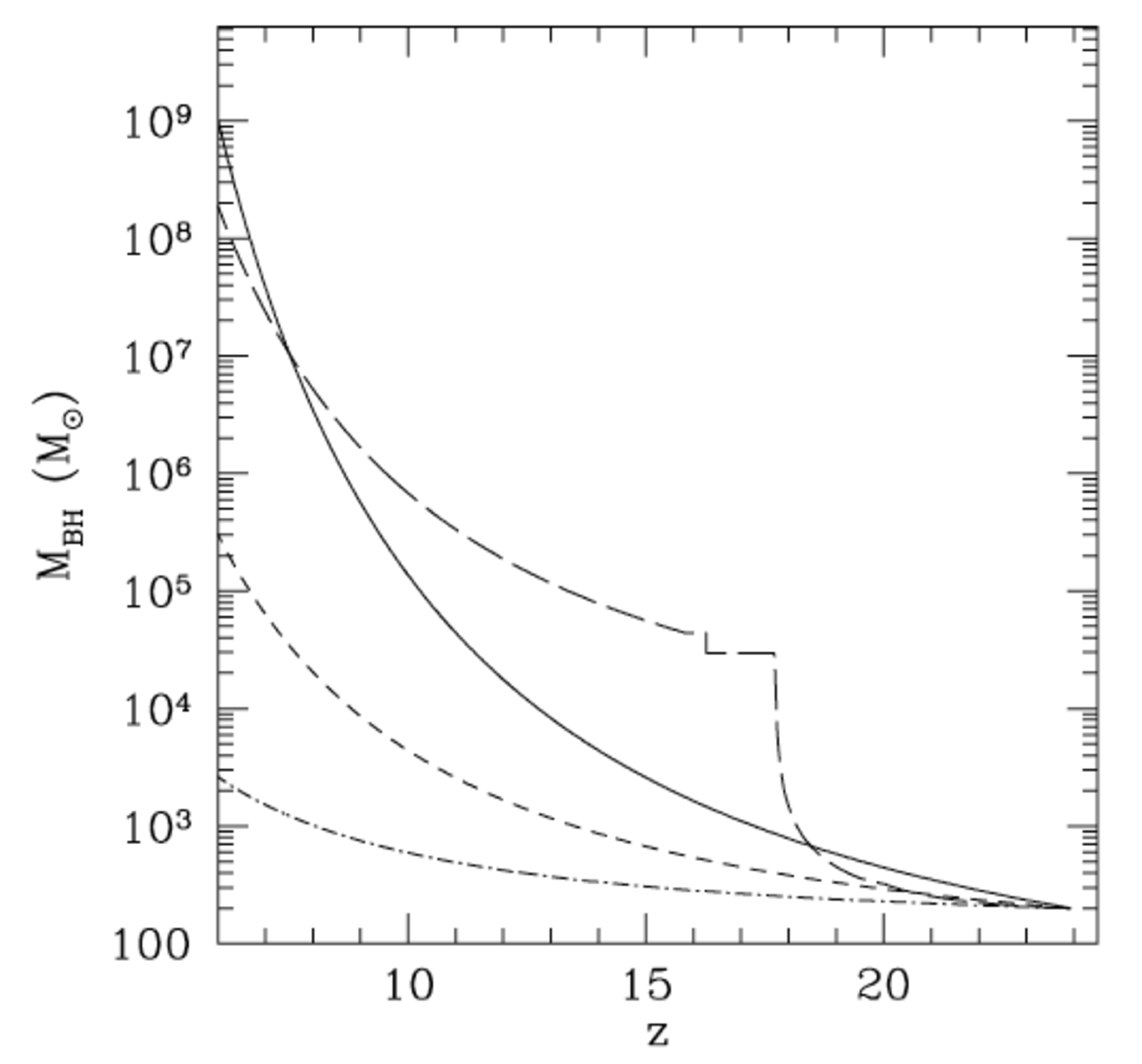}
\caption{The growth of a light seed BH mass as a function of redshift in different regimes: Eddington-limited gas accretion with radiative efficiencies $\epsilon=0.1, 0.2, 0.4$ (solid, short-dashed and dot-dashed lines, respectively); super-critical accretion (long-dashed line). The figure is taken from \citet{VolonteriRees06}. }\label{fig:VR06}. 
\end{center}
\end{figure}

Semi-analytic techniques have been largely employed to study the role of different gas accretion 
regimes and/or the effect of dynamical processes in the early growth of SMBHs (e.g. \citeauthor{Volonteri03} \citeyear{Volonteri03, Volonteri05}; \citealt{Begelman06} \citeauthor{VolonteriRees05} \citeyear{VolonteriRees05, VolonteriRees06}; \citealt{TH09}, \citealt{VSD15}).

\citet{VolonteriRees06} show that the observed high-z SMBH masses can be reproduced starting from light seeds ($\sim 100 \msun$) if they accrete gas at super-Eddington rates, at early stages. 
Super-Eddington accretion is a selective and biased process, occurring only for a small fraction of 
BH seeds if they form in metal-free atomic cooling ($T_{\rm vir}\geq 10^4 \rm K$) halos (e.g.
\citeauthor{VolonteriRees06}, \citeyear{VolonteriRees05, VolonteriRees06}). 

Gas accretion rates that are $10^4$ times higher than the Eddington rate can be reached by light seeds
in super-Eddington models (e.g. \citealt*{VolonteriRees05}, \citealt{P16} and references therein).
However, mildly super-Eddington intermittent accretion at $\sim 3-4 \dot{M}_{\rm Edd}$ (or in general 
$<20 \dot{M}_{\rm Edd}$) may be efficient enough to grow a SMBH in less than 800 Myr (at $z\sim 7$) 
starting from a single (e.g. \citealt{Madau14}, see their Fig. 2) or a population 
(e.g. \citealt{P16}, see their Fig. 5 ) of $100 \msun$ BH seeds. 

In Fig. \ref{fig:VR06} we show the plot presented by \citet*{VolonteriRees06} to illustrate 
the SMBH mass growth along the merger tree of a $10^{13} \msun$ halo at $z=6$. The figure depicts the
effect of different accretion regimes and/or radiative efficiencies on the mass assembly of a 
$\sim 100 \msun$ seed that starts accreting at $z=24$: Eddington-limited with a radiative 
efficiency $\epsilon=0.1, 0.2, 0.4$ (solid, short-dashed and dot-dashed lines, respectively) and
super-Eddington (long-dashed line). Radiatively efficient gas accretion disks ($\epsilon>0.1$) 
strongly limit the growth of their BH, even while accreting continuously at the Eddington rate.

The requirement for episodic, radiatively inefficient, super-critical gas accretion onto stellar mass seed of $20-100\msun$ \, is supported by sub-pc resolution hydrodynamical simulations presented by \cite{Lupi16}. They compare two different methods, the Adaptive Mesh Refinement technique used in the code RAMSES, and the Lagrangian Godunov-type method adopted in GIZMO. 

Super-Eddington gas accretion regimes is not only suitable for light seeds.
In their recent analytic model, \citet{VSD15} show that galactic inflow rates as high 
as $1-100 \rm M_\odot/yr$ may trigger a sequence of fast ($10^4-10^7$ yr) 
episodes of super-critical accretion, onto both light or heavy seeds, at 
rates which are $10^2-10^4$ times larger than in the Eddington-limited 
scenario (see their Fig.~2). 
As a result of these intermittent phases of short super-Eddington 
gas accretion a SMBH can be produced.

In the super-Eddington scenarios, the radiatively inefficient slim disk 
model \citep{Abramowicz88} ensures that even in the presence of 
hyper-Eddington accretion ($>>20\dot{M}_{\rm Edd}$) the bolometric luminosity 
of the accreting BH is only mildly super-Eddington, 
$L_{\rm bol}/L_{\rm Edd}\leq (2-4)$ (e.g. \citealt{Mineshige00}; 
\citealt*{VolonteriRees06}; \citealt{Madau14, VSD15, P16}).

In Eddington-limited gas accretion scenarios, in which the BH can accrete at most at the Eddington rate, the formation of heavy seeds, enabled by the LW radiative feedback is crucial to explain the fast growth of $z\sim 6$ SMBHs (see e.g. \citealt{Johnson13}, the recent pSAMs of \citealt{Petri12, V16} and the review by \citealt{JohnsonHaardt16}). In their mixed-seed-based model \cite{V16} determine the relative contribution of light and heavy seeds to the final BH mass of J1148. 
They report that efficient Eddington-limited growth relies on the formation of 
$\approx 1-10$ heavy seeds in order to produce the expected SMBH mass at $z=6.4$. 
If heavy seed formation is prevented, the predicted final BH mass does not exceed $\sim 10^6$ M$_\odot$, thus warranting the need for super-Eddington accretion in the light
seeds scenario.

Finally, a new cosmological semi-analytic model for galaxy formation, including the growth of SMBHs within a large box size (1.12 cGpc $h^{-1}$) N--body simulation (hSAM), has been presented by \citet{Makiya16}. Their model is currently tuned to reproduce the properties of local galaxies. Using this simulation, \citet{Shirakata16} suggest that stringent constraints on the seed BH mass, may come from less massive bulges observed at $z\sim 0$, rather than the high redshift BH-bulge mass relation. Their study suggests that the mass of BHs observed in $\sim 10^9 \msun$ bulges is overpredicted if only seeding by heavy seeds ($10^5 \msun$) is considered. Such small stellar mass bulges instead favour seeding by smaller seed BHs ($10^3 \msun$) or a mixed population of seed BHs randomly distributed in the mass range $10^3-10^5 \msun$.


A suite of high spatial resolution simulations ($\sim 10$ pc) have been devoted to study the effect of galaxy mergers on BH accretion, as a function of the initial merging galaxies' mass ratio, orbital configuration and gas fraction. These different stages of galactic encounters is described in \citet{Capelo15}. They confirm that more efficient BH accretion is induced during galaxy mergers with the initial mass ratio being the most critical parameter affecting BH accretion and AGN activity.

In the simulations presented by \citet{Feng14, DiMatteo16} the rapid growth of BHs, occurring in bulge dominated galaxies, is driven by large scale filamentary cold gas accretion, rather than by major gas rich mergers. \citet{Feng2014} extract 3 DM halos from the cosmological hydrodynamical simulation MassiveBlack, hosting $10^9 \msun$ BHs and re-simulate them with zoom-in techniques. They find that dense cold gas is able to sustain accretion. During the accretion phase at the Eddington rate, the cold gas directly feeds the BH, while in the sub-Eddington phase (that they find for $z \lesssim 6$), the accretion disc is disturbed and disrupted by feedback.

\subsection{BH feedback}
\label{sec:feedback}

As discussed in section \ref{sec:questions}, the physical processes involved in quasar 
formation and evolution are expected to be regulated by AGN and stellar feedback.
During the quasar-dominated regime ($z\lesssim 8$, see section \ref{sec:coevo}) a strong, 
galaxy-scale wind is predicted to be driven by the energy released during both BH accretion 
and SN explosions. 
This feedback is expected to clear the ISM of gas and dust leaving a un-obscured 
line of sight toward the central emitting source. 
In addition, radiation emitted from the optically bright quasar J1148 may contribute to at least $30\%$ of the observed FIR luminosity ($>20\ \mu$m) heating the large amount of dust ($\sim 3\times 10^8\msun$) in the host galaxy ISM, outside the un-obscured cone. Both stellar and quasar optical/UV emission are expected to be reprocessed by dust, thus contributing to the observed FIR luminosity \citep{Schneider15}.

Adopting an energy-driven wind prescription similar to that usually adopted by numerical simulations (e.g. \citealt{DiMatteo05}) pSAMs show that the AGN feedback is the main driver of the massive observed gas outflow rates at $z>6$. This is predicted to have a dominant effect with respect to stellar feedback (energy-driven winds from SN explosions) in shaping the high-z BH-host galaxy co-evolutionary path. For example, a powerful quasar-driven gas outflow is launched during the latest stages of the evolution ($\sim 100-200$ Myr) in the best-fit models of \citeauthor{V11}, \citeyear{V11, V12} and \citet{P16}, for J1148. The predicted outflow rates are in good agreement with the observations, $>1000-3000$ M$_\odot/$yr \citep{Maiolino12, Cicone15} and $\sim 10^3$ times more efficient than the sub-dominant SN-driven contribution.  

However, it is worth noting that the prescription usually adopted in SAMs to describe the energy-driven wind effects can not provide insights on the physical processes determining the observed properties of the outflowing gas and its complex dynamics.
 
Although described by sub-grid prescriptions, the response of the gas to the energy released by the accreting BH is now well described by hydro-dynamical simulations.
\citet{Costa2014} study AGN feedback using the moving-mesh code AREPO. 
They find that, despite the fact that momentum driven outflows 
predict a $M_{\rm BH}-\sigma$ relation similar to that observed, 
the energy-driven scenario better reproduces the observed, large scale anisotropic
AGN-driven outflows. 
With the same code \citet{Costa2015} re-simulate a zoom-in region around the six most 
massive halos at $z \sim 6$ to study the brightest quasars. 
They show that the high-velocity extended cold gas observed out to $\sim 30$ kpc 
(\citealt{Maiolino12, Cicone15}) requires the combined effect of SN and AGN feedback. 
SN-driven winds are responsible for the pre-enrichment of the circumgalactic
and intergalactic medium in which the massive, fast ($>1400\ \rm km s^{-1}$) AGN-driven 
hot outflow is launched, ensuring efficient radiative cooling (see e.g. Fig.~2 in 
\citealt{Costa2015}) to explain the presence of cold gas (see e.g. \citealt{Cicone15}).

Finally, high velocity ($10^2-10^3$ km s$^{-1}$) energy-driven winds
on large scales have been recently also studied by \citet{Bieri16}
by means of radiation-hydrodynamic simulations of isolated galactic discs.
They suggest that outflow rates as high as $\sim 10^3$ M$_\odot/yr$ are sustained by IR radiation, with 
scattering on dust grains enabling efficient momentum transfer to the gas.

\section{The host galaxy properties}
\label{sec:hosts}
\subsection{The origin of high-z dust.}

Several theoretical models have been devoted to the study of the 
rapid enrichment of the ISM in $z>6$ galaxies and quasars, and 
in particular to the origin of the huge amount of dust ($>10^8\msun$)
inferred from the FIR and sub-mm observations 
(e.g. \citealt*{HF02, ME03, DGJ07, DC11}; 
\citeauthor{V09}, \citeyear{V09, V11, V14};
\citeauthor{Gall11a}, \citeyear{Gall11a, Gall11b}; 
\citealt{Mattsson11}, \citealt{V14, Calura14}).

A SN origin for the dust observed in the early Universe has often 
been advocated because of the shorter evolutionary
time scale of core collapse SNe progenitors ($10-40\msun$ stars, 
with an age $<10$ Myr) with respect to that of AGB stars 
(e.g. \citealt*{ME03, DGJ07}).
This scenario was supported by the deviation of the dust extinction curves of 
$z>4$ quasars and gamma ray bursts (GRB) from the Small
Magellanic Cloud (SMC) extinction curve, typical of $z<2$ quasars 
\citep{Maiolino04, Stratta07, Perley10, Gallerani10}. 
This suggests either a different dust production mechanism or 
dust processing into the ISM at high redshift. 

However, subsequent studies pointed out that stellar sources alone 
can not account for the entire dust budget and grain growth in cold, 
dense gas clouds must also have a dominant role, even at $z>6$ (e.g. \citealt{Michalowski10, 
V11, Pipino11, Rowlands14}; but see \citealt{Ferrara16}).

Moreover, in contrast to what was previously thought, AGB stars are able to 
significantly contribute to dust production in high redshift quasars, 
producing a dust mass at least similar to that of SNe, already at $z\sim 8-10$ 
depending on the host galaxies' SFH and IMF (see \citealt{V09} and Fig.~8 in 
\citealt{V11}). 

Modeling the properties, and in particular the evolution of dust, in quasar 
host galaxies at $z>6$ is still a major challenge.
\citeauthor{Li07} \citeyear{Li07, Li08} carried out the first multi-scale simulation, 
using GADGET2 \citep{Springel05}, 
aimed to follow the formation of quasar J1148 in a hierarchical scenario, 
accounting for self-regulated BH growth (starting from Pop~III seeds), AGN feedback 
and the host galaxy properties evolution.
They showed that the metallicity and dust mass of J1148 are produced through a series of 
efficient bursts of star formation (see Fig.~7 in \citealt{Li07}) resulting in a final 
stellar mass of $10^{12} \msun$, similar to what is expected from the local 
$M_{\rm BH}-M_{\star}$ relation. To date, this is the only attempt to study the high-z dust
properties made with numerical approaches \citep{Li08}.
However, only a single plausible hierarchical build-up of the J1148 
DM halo, extracted (and re-simulated) from the $1 h^{-1}$ Gpc$^3$ volume is explored
in these works and thus, the resulting SFH is unique.
Semi-analytic models, which instead enable a statistical investigation of different SFHs, 
provide similar conclusions. The chemical properties of the host galaxy require an order 
of magnitude higher stellar mass with respect to the dynamical constraint, as discussed in the following sections.

\subsection{The BH-host galaxy co-evolution}
\label{sec:coevo}
\begin{figure*}
\begin{center}
\includegraphics[width=5.5cm]{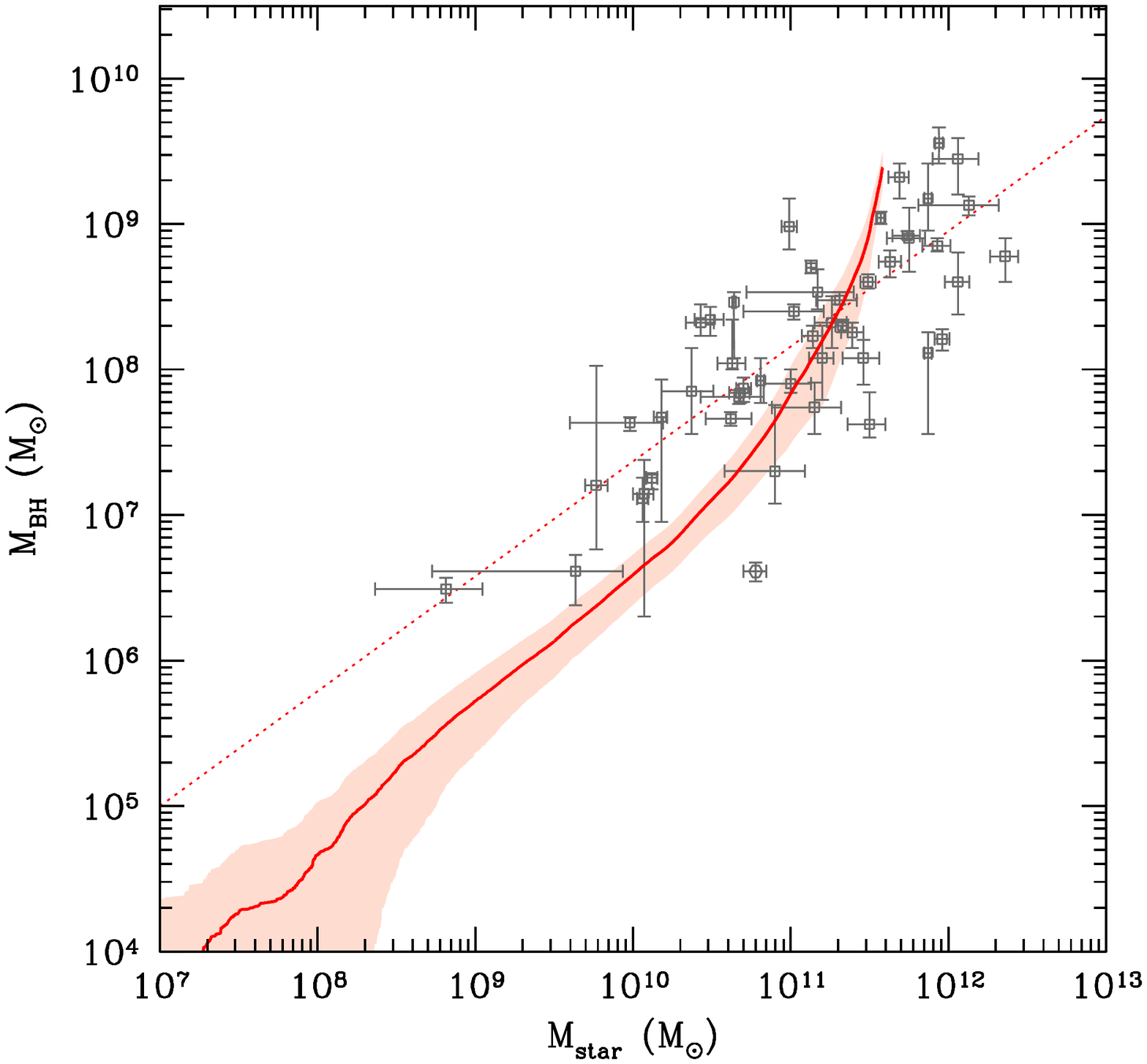}
\includegraphics[width=5.5cm]{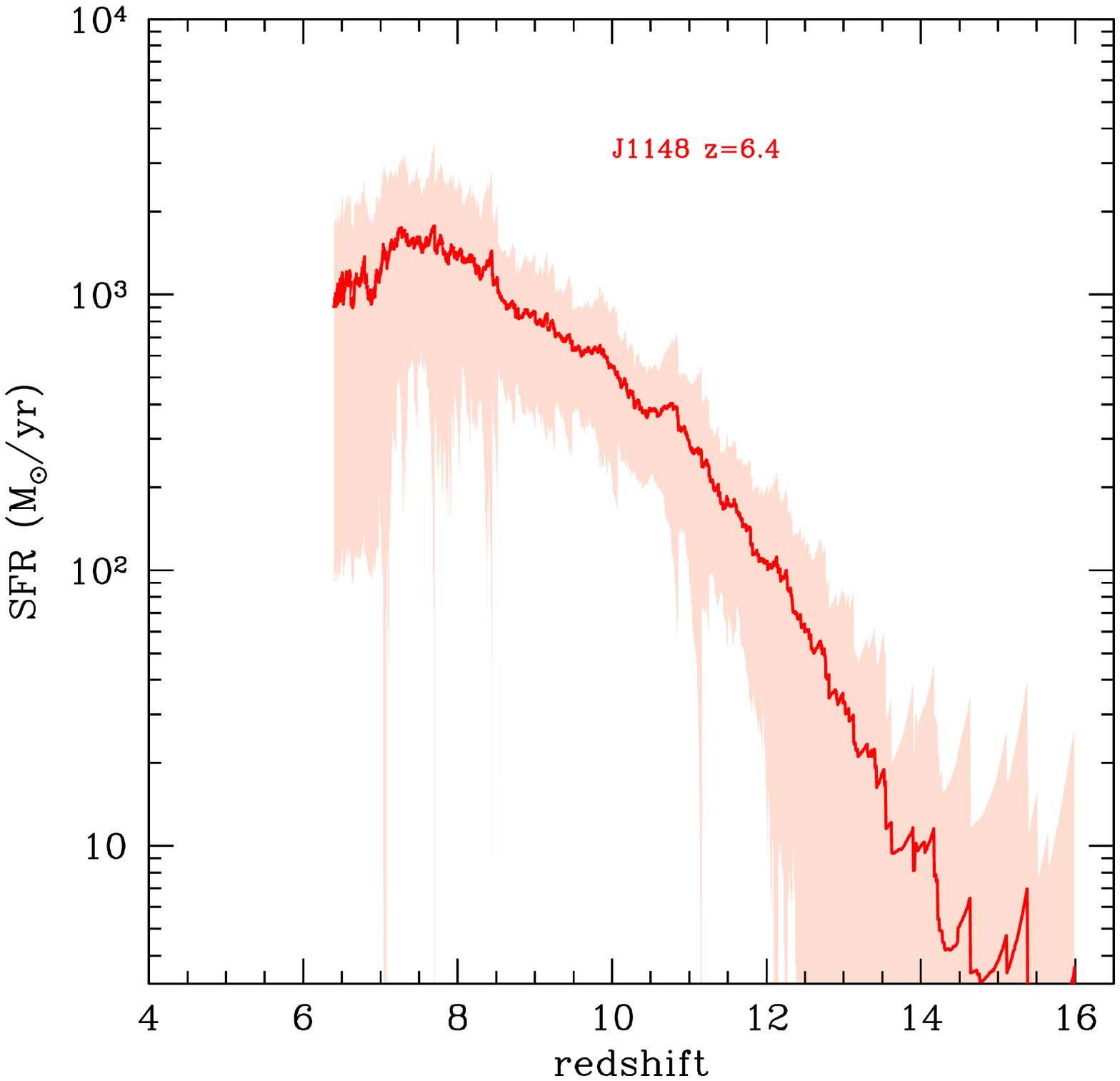}
\includegraphics[width=5.5cm]{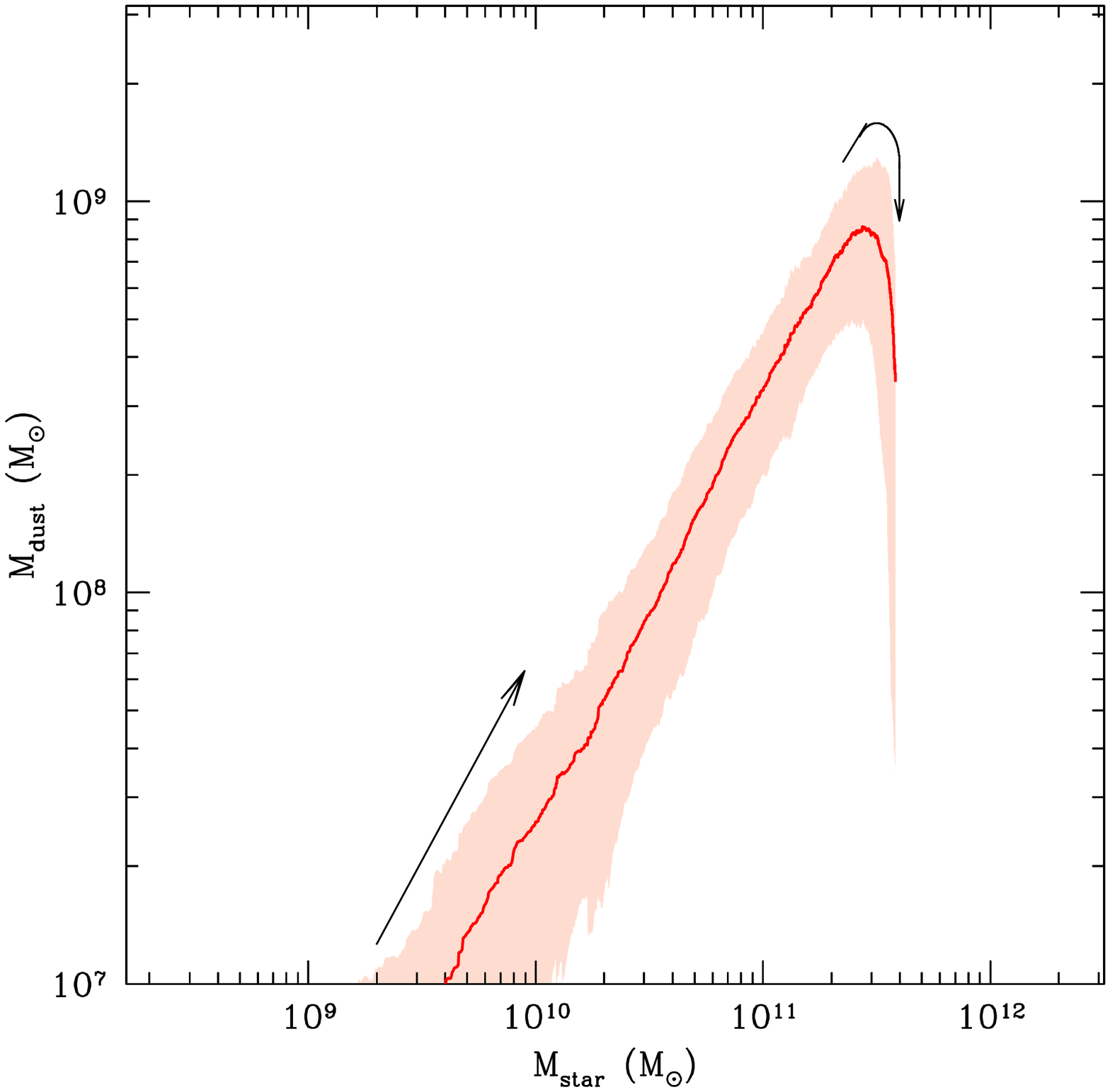}
\caption{The cosmic cycle of a typical quasars at $z\sim 6$. Models reproduce the properties
of J1148 (see text).
Left panel: the build-up of the $M_{BH}-M_{star}$ relation through cosmic time as compared
with data and empirical fit for local galaxies \citep{Sani11}. Middle panel: the 
predicted star formation history via quiescent and merger-driven bursts (see e.g. 
\citealt{V11}). Left panel: the assembly of the dust mass into the ISM as a function of the 
stellar mass. In all panels the solid lines show the average over 50 different DM halo 
merger trees with shades representing the $1\sigma$ dispersion. These figures are adapted from
\citet{V11}. }\label{fig:CosmicCycle}
\end{center}
\end{figure*}
Observational campaigns at $z>5$ show that quasars and their host galaxies 
are characterized by similar properties in terms of the BH, dynamical, dust and 
molecular gas masses, suggesting a common evolutionary scenario.

In Fig.~\ref{fig:CosmicCycle} we show the best-fit evolutionary scenario for the
BH and host galaxy properties of J1148 as predicted by \citeauthor{V11} (\citeyear{V11, V14}).
Solid lines show the redshift evolution of the total masses\footnote{At each redshift the total BH 
mass is given by the sum of the masses of all the existing nuclear BHs. In the same way the
total stellar and dust masses represent the stellar and dust content summed over all the existing
halos. See \citet{V11} for details.} of BH and stars (on the left), the total 
SFR (in the middle) and dust and stars again (on the right) averaged over 50 
different DM halo merger trees, with shaded areas representing the $1\sigma$ 
error.

As soon as efficient star formation starts, the BH grows in the buried AGN. At this stage its 
optical emission is outshined by the ongoing strong star burst, SFRs from 100 up to $>1000$ M$_\odot/$ yr, at $z\sim 8$ (middle panel). The mass of dust (right panel) rapidly grows, reaching values as high
as $10^9\msun$, when the bulk of the stellar mass, $\sim (2-4)\times 10^{11}\msun$, is already in place. During this dust-obscured phase, the total nuclear BH mass reaches $\sim 2\times 10^8\msun$.

In this scenario, the progenitor galaxies of J1148 at $z\sim 8-10$ are 
predicted to have similar properties (e.g. BH, stellar and dust mass) as the observed sub-mm galaxies (SMGs) at lower redshifts
(e.g. \citealt{Santini10, Michalowski10, Magnelli12}). These sub-mm galaxies are suggested to
be the evolutionary stage preceding the active quasar phase.

The transition between the starburst-dominated regime and the quasar-dominated evolution, at $z<8$,
is triggered by powerful energy-driven winds which clear up the ISM of dust and gas (see 
e.g. the down turn indicated by the black arrow in the right panel of Fig~\ref{fig:CosmicCycle}),
un-obscuring the line of sight toward the quasar and damping the SFR (we will discuss the AGN feedback in the following section). 

SMBH evolution models suggest a steeper evolution of the BH-stellar bulge mass 
relation at high redshift, with the SMBH forming before/faster than 
the stellar bulge (e.g. \citealt{Petri12}). In addition, the observed deviation 
of high redshift quasars from the local BH-stellar-mass ratio seems to be a natural outcome of SMBH growth driven by episodic super-Eddington 
accretion which leads to a BH accretion rate-to-SFR ratio of $>10^2$ \citep{VSD15}.\\
\citet{Agarwal13} track the subsequent growth of DCBH seeds by using a modified version of the \citet{Agarwal12} hSAM. In their simulated volume, they find that the merger of a DCBH host satellite with the neighboring galaxy (source of the LW radiation field), leads to the resultant system lying above the local M$_{\rm BH}$--M$_{\rm star}$ relation, already at these early stages of the evolution. The authors term this phase as `obese black hole galaxies' or OBGs as the DCBH is able to outshine the stellar component, leading to unique observables that distinguish them from normal galaxies. The OBGs are expected to transition onto the local M$_{\rm BH}$--M$_{\rm star}$ relation via mergers.
However, they do not account for the formation and evolution of metals and dust in 
the ISM, which represent a strong constraint on the host galaxy SFH and final
stellar mass.

Chemical evolution models instead point out that SFR, gas, metals and dust 
content of quasar host galaxies are well reproduced with standard assumptions 
of stellar initial mass function (IMF), star formation efficiency and dust 
grain growth, for galaxies with stellar masses $\geq 10^{11}\msun$ (see left panel of
figure \ref{fig:CosmicCycle}). 
These are about one order of magnitude higher than the stellar masses inferred 
from the observations of high redshift quasars (e.g. \citeauthor{Wang10} 
\citeyear{Wang10, Wang13}) and would bring the predicted $M_{\rm BH}-M_{\rm star}$
relation closer to the local value, suggesting that high redshift dynamical 
(and thus stellar) masses may be underestimated (\citeauthor{V11}, 
\citeyear{V11, V14}, \citealt{Calura14}).

Although a top-heavy IMF scenario (i.e. biased to more massive stars) {can} 
reproduce the observed dust mass and the deviation of J1148 from the local 
$M_{\rm BH}-M_{\rm star}$ relation, it requires a {less-efficient SFH to do so. This results in a 
SFR at $z=6.4$ that is more than 10 times smaller than the observed rate (\citealt{V11}), too small to provide the observed FIR luminosity even if the AGN contribution to dust heating \citep{Schneider15} is accounted for.}

{Instead, assuming a short evolutionary time scale does not solve the tension either.
At the observed SFR $\sim 1000$ M$_\odot/$yr the $\sim (3-4)\times 10^{10}\msun$ stellar mass estimated for quasars like J1148 would be produced in 
a quite short time interval, $\sim 10-20$ Myr. Such an evolutionary time scale 
is too short for stellar evolution to account for dust enrichment up to
 $>10^8\msun$, even with a maximally efficient mode of dust formation by SNe 
(see \citealt{V14} for a detailed discussion).}

Following this discussion, it is important to note that, at $z>6$, stellar 
masses can not be convincingly obtained via SED fitting as 
in local and lower redshift systems. A lower limit to the stellar mass 
(dynamical bulge) is usually obtained as $M_{\rm star}=M_{\rm dyn}-M_{\rm H_2}$ where 
dynamical and molecular gas masses, $M_{\rm dyn}$ and $M_{\rm H_2}$, 
respectively, are derived from CO observations. 

Large uncertainties are introduced by the methods adopted to infer 
$M_{\rm dyn}$ and $M_{\rm H_2}$. A large scatter ($>60\%$) in the estimated 
molecular gas mass is due to the adopted CO line luminosity$-$to$-$H$_2$ mass 
conversion factor, $\alpha_{\rm CO}=0.8 \pm 0.5$ M$_\odot/$(K km s$^{-1}$ pc$^2$).
This is typical of ultra luminous infrared galaxies (ULIRGs, 
\citealt{Solomon97}, \citealt*{DownesSolomon98}) and usually adopted for high 
redshift quasars too.
In addition, $M_{\rm dyn}$ strongly depends on geometrical assumptions for 
gas distribution which is usually considered to be disk-like, with given 
inclination angle $i$ and radius $R$, which are 
difficult to infer from observations at such high redshifts.
An uncertainty of more than $50\%$ must be associated to the inferred values,
$M_{\rm dyn}sin^2i = (10^{10}-10^{11})\msun$.
A radius $R=2.5$ kpc and an inclination angle $i=65°$ have been inferred for 
J1148, in which the CO emitting region is spatially resolved \citep{Walter04}. 
For other quasars a similar radius and a mean inclination angle of $40°$ are 
usually assumed (see e.g. \citealt{Maiolino07, Wang10}). 

Theoretical studies suggest that lower inclination angle ($i<30°$) and/or 
larger disk radius ($R\sim 5-30$) kpc may solve the so-called stellar mass 
crisis (see e.g. Fig.~9 and discussion in \citealt{V14}).

Recent Atacama Large Millimeter and sub-millimeter Array (ALMA) observations 
of [CII] emission in quasars have suggested that a large fraction of the CO may 
be still undetected \citep{Wang13}, supporting the idea that dynamical mass 
estimates could be missing some of the stars.
Moreover, IRAM Plateau de Bure Interferometer (PdBI) follow-up 
observations of [CII] $158 \mu$m emission line and FIR continuum in J1148 host 
galaxy have revealed the presence of an extended cold gas component out to 
$\sim 30$ kpc which may be an indication of star formation activity on 
larger scales with respect to the size of the CO emission \citep{Cicone15}. 

Thus, stellar mass estimates from model predictions and observations may
be reconciled by accounting for a more complex and/or more extended star and gas distribution, 
beyond the few kpc radius inferred from the CO emitting regions.
Observations \citep{Cicone15}, SAMs (\citeauthor{V11}, \citeyear{V11, V14} 
and \citealt{Calura14}) and numerical simulations (e.g. \citealt{Khandai12}) 
seem to agree with this scenario.
Quasars at $z\sim5$ resolved in the MassiveBlack simulation are predicted to be 
compact and gas rich systems with intense burst of star formation occurring in 
both the innermost and outer regions, out to the DM halo virial radius
($\sim 200 h^{-1}$ kpc, \citealt{Khandai12}).

In addition, \citet{DiMatteo16} show that the most massive 
BHs ($>10^8\msun$) at $z\sim 8$ reside in compact bulge-dominated galaxies 
(more than $80\%$ of the stars are in the spheroidal component). 
The total stellar masses of these systems are already $>10^{10}\msun$ (see 
e.g. Fig~1 and Table 1 of \citealt{DiMatteo16}), bringing them well within the 
scatter of the observed local $M_{\rm BH}-M_{\rm star}$ relation.
Pure SAMs provide very similar results \footnote{A mean BH and stellar mass of $4\times 10^8$ and $3-4\times 10^{10}\msun$ are predicted in both light- and mixed-seeds scenarios presented in \citet{V16, P16}.}. 

\section{Discussion}
\label{sec:discussion}

In this review we have discussed the formation of the first quasars, and in particular the rapid growth of their SMBHs focusing on pure semi-analytic or hybrid (SAM plus N--body simulations) approaches.

For comparison, we have also mentioned the results of some of the state-of-the-art hydrodynamical simulations, providing deep insights on the dynamical evolution of galaxies. With respect to these simulations, semi-analytic (pure or hybrid) methods have the complementary role of enabling statistical studies and exploring different models and parameter space, on shorter computational time scales. 

However, simplified geometries, models for the gas cycling and/or sub-grid prescriptions limit the scope of both pSAMs and hSAMs. Indeed, some physical aspects are still far from being taken into account in these models, such as the gas physics, feedback from stars and/or the accreting BH, or accretion rate in the inner part of the halo. This is where cosmological hydrodynamical simulations offer a laboratory to study the impact of physical processes related to the structure of collapsed objects. \\
Angular momentum, for example, is one such physical process. Gravitational systems, such as halos can possess a given degree of rotational support, which is described by the spin parameter $\rm{\lambda_{spin}}=J\vert E \vert^{1/2}/GM_{\rm h}^{5/2}$, with $J$ the angular momentum of halos, $E$ the total energy, and $M_{\rm h}$ the mass of halos. The angular momentum of a halo, or its baryonic central region, is thought to be the result of clustering/surrounding neighbors applying tidal torques on the given halo \citep{Peebles69}.


Although, they have the advantage of directly tracking the cosmic evolution of the baryonic component of galaxies (where semi-analytic models need to use approximations), the main limitation of hydrodynamical simulations is that the physical processes, acting on different scales can not be described simultaneously, yet\footnote{In addition, due to the higher computational costs required to run hydrodynamical simulations these models are often restricted to few realizations, small volumes and/or still require sub-grid prescriptions (just like SAMs).}. In other words, large and small scales can not be resolved at the same time in simulations. This has been widely discussed by \citet{Habouzit16hydro}, in the case of DCBH formation. They use a small scale (1 cMpc), high resolution ($M_{\rm DM, res}\sim 2\times 10^3\msun$) to study in detail the effect of expanding metal-rich bubbles around possible DC sites, while a larger box size (10 cMpc) with intermediate resolution ($M_{\rm DM,res}\sim 10^7\msun$) is adopted to statistically asses the impact of metal enrichment, SFR and SN-driven winds on the DCBH number density, in a significant volume of the Universe. Finally, the Horizon-noAGN large box (142 cMpc), low resolution ($8\times 10^7\msun$) simulation is adopted to test whether DCBHs are able to explain the population of high redshift quasars.  

Among the most recent hydrodynamical simulations devoted to study the rare, high density peaks DM halo hosting the first quasars, MassiveBlack \citep{DiMatteo12} and its high-resolutions zooms \citep{Khandai12, Feng14}, investigate the formation of SMBHs in the first galaxies, by covering a volume of 0.75 Gpc$^3$.
A higher resolution is reached in the 0.5 Gpc$^3$ volume of the BlueTides simulation (\citeauthor{Feng15} \citeyear{Feng15, Feng16}), enabling the study of the formation of the first SMBHs at early cosmic epochs ($z>7$, \citealt{DiMatteo16}).

Given the advancement in theoretical modeling techniques, all the different approaches can together be considered as a powerful tool to investigate different physical processes related to the formation and evolution of the first quasars at $z\sim 6$. Combined with observational constraints from current and future high-resolution instruments, these models can be further improved to provide definitive answers to the open questions discussed in section \ref{sec:questions}.

\begin{acknowledgements}
We thank Raffaella Schneider and Marta Volonteri for triggering this review.
RV thanks the EWASS2015 organizers for including the Symposium on
\textit{Understanding the growth of the first supermassive black holes}
in their program. RV and EP thank F. Weng and X.~B. Wu for providing useful data. 
M.H. thanks M. Latif and M. Volonteri for useful discussion. BA would like to acknowledge the funding from the European Research Council under the European Community?s Seventh Framework Programme (FP7/2007-2013) via the ERC Advanced Grant STARLIGHT (project number 339177).
RV has received funding from the European Research Council under the European 
Unions Seventh Framework Programme (FP/2007-2013) / ERC Grant Agreement n. 306476
while writing the manuscript.
\end{acknowledgements}

\bibliographystyle{pasa-mnras}
\bibliography{biblio_rev}

\end{document}